\documentclass[numbers]{frontiersSCNS}

\usepackage[bookmarks]{hyperref}  %
\usepackage[margin=.7in]{geometry}
\usepackage{amsmath}
\usepackage{amsfonts}
\usepackage{xcolor}
\usepackage{bm}  %

\renewcommand{\vec}[1]{\bm{#1}}
\newcommand{\mat}[1]{{\mathbf{#1}}}
\newcommand{\diff}{\mathrm{d}}  %
\newcommand{\defeq}{\stackrel{\mathrm{def}}{=} }  %

\newcommand{\degree}{\kappa}
\newcommand{\numNodes}{{N}}
\newcommand{\couplingK}{{K}}
\newcommand{\numCofs}{{M}}
\newcommand{\basisFuncSymbol}{\psi}
\newcommand{\basisFunc}[1]{\basisFuncSymbol^{(#1)}}
\newcommand{\basisFuncSubscript}[2]{\basisFuncSymbol^{(#1)}_{#2}}
\newcommand{\trueEig}{\lambda}
\newcommand{\coarseApproxdEig}{{\hat{\lambda}}}
\newcommand{\coarseEig}{{\mu}}
\newcommand{\avgPhase}{{\bar{\theta}}}
\newcommand{\SI}{r}

\newcommand{\xMeasure}{\Gamma}
\newcommand{\diffMeasureVec}{{\diff \xMeasure(\vec x)}}
\newcommand{\intD}{{\int_D}}
\newcommand{\numSamps}{{n_\text{samp}}}

\newcommand{\figRef}[1]{Fig. \ref{fig:#1}}
\newcommand{\eqnRef}[1]{Eq. (\ref{eqn:#1})}
\newcommand{\secRef}[1]{\S\ref{sec:#1}}
\newcommand{\appendixRef}[1]{Appendix \ref{sec:#1}}

\newcommand{\omegaOneDFunc}{\xi}
\newcommand{\degreeOneDFunc}{\zeta}
\newcommand{\multiIndexCharacter}{\gamma}
\newcommand{\multiIndex}{\vec{\multiIndexCharacter}}
\newcommand{\indexSet}{G}
\newcommand{\scalarFuncIndex}{k}
\newcommand{\scalarFuncIndexAlt}{l}

\newcommand{\multiIndexA}{\multiIndexCharacter_{\scalarFuncIndex,\omega}}
\newcommand{\multiIndexB}{\multiIndexCharacter_{\scalarFuncIndex,\degree}}

\DeclareRobustCommand{\inTstep}{\tau}
\DeclareRobustCommand{\outTstep}{h}

\newcommand{\wordDefn}[1]{\emph{#1}}

\usepackage{url,hyperref,lineno,microtype,subcaption}
\usepackage[onehalfspacing]{setspace}

\def\keyFont{\fontsize{8}{11}\helveticabold }
\def\firstAuthorLast{Bertalan {et~al.}} %
\def\Authors{
Tom Bertalan\,$^{1}$,
Yan Wu\,$^{1,2}$,
Carlo Laing\,$^{3}$,
C. William Gear\,$^{1}$,
and
Ioannis G. Kevrekidis\,$^{1,4,*}$
}
\def\Address{
$^{1}$Department of Chemical and Biological Engineering, Princeton University, Princeton, New Jersey, USA\\
$^{2}$Department of Bioengineering, University of California, La Jolla, California, USA\\
$^{3}$Institute of Natural and Mathematical Sciences, Massey University, Auckland, New Zealand\\
$^{4}$Program in Applied and Computational Mathematics, Princeton University, Princeton, New Jersey, USA
}
\def\corrAuthor{Ioannis G. Kevrekidis,
Program in Applied and Computational Mathematics,
Fine Hall,
Washington Road, Princeton NJ
08544-1000 USA
}

\def\corrEmail{yannis@princeton.edu}

\usepackage{graphicx} %
\usepackage{mathtools}  %

\newcommand{\chungLuPVal}{0.50}
\newcommand{\chungLuQVal}{0.90}
\newcommand{\chungLuRVal}{0.50}
\newcommand{\baseNVal}{196}
\newcommand{\numReplicatesVal}{32}
\newcommand{\numCofsVal}{28}
\newcommand{\defaultKVal}{1}
\newcommand{\absTolPowVal}{-6.0}
\newcommand{\errBarTypeVal}{$\pm$ 1 standard deviation}

\newcommand{\bigBigNVal}{10000}
\newcommand{\plainScatterKVal}{0.5}
\newcommand{\KuramotoCPItStepVal}{0.45}
\newcommand{\KuramotoCPIatollogTenVal}{-6.0}
\newcommand{\KuramotoCPIrtollogTenVal}{-12.0}
\newcommand{\dhdtDeltaTVal}{0.417}
\newcommand{\dhdtrtolLogTen}{-12.0}
\newcommand{\dhdtatolLogTen}{-12.0}
\newcommand{\cstrVVal}{0.4}
\newcommand{\cstraVal}{2}
\newcommand{\cstrbVal}{1}
\newcommand{\cstrxAinVal}{1}
\newcommand{\cstrxBinVal}{0}
\newcommand{\cstrxAInitVal}{1}
\newcommand{\cstrxBInitVal}{0}
\newcommand{\flowTimeVal}{0.05}
\newcommand{\pseudoArclengthTauVal}{0.30}
\newcommand{\eigMethodThreshVal}{3}
\newcommand{\fdJacDiffVal}{0.001}
\newcommand{\hiOmVal}{0.100}
\newcommand{\omStdVal}{0.060}
\newcommand{\eigcompsNval}{196}
\newcommand{\eigcompsMval}{28}
\newcommand{\graphPlotNval}{4000}
\newcommand{\cpiNval}{300}
\newcommand{\degreeSliceRadiusVal}{100}

\begin{document}
\onecolumn
\firstpage{1}

\title[Coarse-Grained Descriptions with Structural Heterogeneities]{
    Coarse-Grained Descriptions of Dynamics for Networks with both Intrinsic and Structural Heterogeneities
}

\author[\firstAuthorLast ]{\Authors} %
\address{} %
\correspondance{} %

\extraAuth{}%

\maketitle

\begin{abstract}
Finding accurate reduced descriptions for large, complex, dynamically evolving
networks is a crucial enabler to their simulation, analysis, and, ultimately, design.
Here we propose and illustrate a systematic and powerful approach
to obtaining good collective coarse-grained observables--variables 
successfully summarizing the detailed state of such networks.
Finding such
variables can naturally lead to successful reduced dynamic models for the
networks.
The main premise enabling our approach is the assumption that the behavior of
a node in the network depends (after a short initial transient) on the \wordDefn{node identity}:
a set of descriptors that quantify the node properties, whether intrinsic (e.g. parameters
in the node evolution equations) or \wordDefn{structural} (imparted to the node by its connectivity 
in the particular network structure).
The approach creates a natural link with modeling and ``computational enabling technology" developed
in the context of Uncertainty Quantification.
In our case, however, we will not focus on ensembles of {\em different} realizations
of a problem, each with parameters randomly selected from a distribution.
We will instead study many {\em coupled heterogeneous} units, each characterized by 
randomly assigned (heterogeneous) parameter value(s).
One could then coin the term \wordDefn{Heterogeneity Quantification} for this approach,
which we illustrate through a model dynamic network consisting of coupled oscillators
with one {\em intrinsic} heterogeneity (oscillator individual frequency) and one {\em structural} heterogeneity 
(oscillator degree in the undirected network).
The computational implementation of the approach, its shortcomings and possible extensions are also discussed.
\tiny
 \keyFont{ \section{Keywords:}
    nonlinear dynamics, heterogeneity, instability, uncertainty, bifurcation,
    data mining, machine learning, networks, model reduction.
 }
\end{abstract}

\section{Introduction}
\label{sec:introduction}

Model reduction for dynamical systems has been an important
research direction for decades; accurate reduced models are very useful,
and often indispensable for the understanding, analysis, and ultimately
for the design of large/complex dynamical systems.
The relevant tools and techniques range from center manifold reduction close to
bifurcation points \cite{Guckenheimer2002}
to singular perturbation techniques (analytical \cite{Bender2015} or computational
\cite{Kevorkian1996}%
)
and (more recently) to data-driven reduction methods (like PCA \cite{Jolliffe2002},
or nonlinear manifold learning techniques \cite{Coifman2005,Dsilva2016}).
While many such tools are well established for ODEs and PDEs, the dynamics of networked dynamical
systems (e.g. \cite{Newman2006}) pose additional challenges.

We are interested here in large sets of dynamic units (agents, oscillators, cells) linked in
a prescribed (and, for this paper, fixed) coupling pattern.
Every unit consists here of a (relatively small) set of ordinary differential equations.
The units are \emph{intrinsically} heterogeneous, meaning that the parameters of this set of ODEs are sampled
from a probability distribution.
Once the overall system of ODEs
modeling a large network is assembled, \emph{any} generic dynamical system model reduction technique can be tried.
For all-to-all coupled heterogeneous units, in particular, there has been extensive
work taking advantage of the overall model structure, leading to the systematic reduction of such
intrinsically heterogeneous assemblies \cite{Laing2016,Ott2008,Ott2009}.

Our illustrative example is a simulation of coupled phase oscillators
whose dynamics are governed by the equations
\begin{equation}
\label{eqn:kuramoto}
    \frac{\diff \varphi_i(t)}{\diff t} = \hat\omega_i +
    \frac{\couplingK}{\numNodes}\sum_{j=1}^\numNodes A_{i,j}\sin(\varphi_j(t) - \varphi_i(t)),
\end{equation}
where $i\in1,\ldots,\numNodes$,
and
$A_{i,j}\in\{0,1\}$ is the adjacency matrix for a network with identical edges.
This model was originally formulated by Yoshiki Kuramoto with all-to-all coupling
(i.e., $A_{i,j}=1\forall i,j$)
\cite{Kuramoto1975,Kuramoto1984}.
While we work with the simplified Kuramoto oscillator system,
the methods presented here have been shown to work
for more realistic coupled-oscillator systems as well,
such as Hodgkin-Huxley-like neurons \cite{Choi2016},
metabolizing cells \cite{Bold2007a},
gene-expression oscillations in circadian rhythms (ongoing work),
or other candidate systems \cite{Ashwin2016}.
In cases such as the Hodgkin-Huxley, where each unit is described by multiple dynamic variables,
(e.g, membrane potential and gating variables),
the analysis used here is repeated for each per-cell variable.

In order to construct a frame that moves with the average phase angle,
we use states $\vec{\theta} \in \mathbb R^{\numNodes-1}$,
where
\begin{equation}
 \label{eqn:thetaDefinition}
\theta_i(t) \defeq
\left(
    \varphi_i(t) - \frac{1}{\numNodes} \sum_{j=1}^{\numNodes} \varphi_j(t)
\right),
\quad\quad
i = 1,2, \ldots, \numNodes-1,
\end{equation}
and
$\theta_\numNodes(t) \defeq -\sum_{j=1}^{\numNodes-1} \theta_j(t)$.
(Hereafter, explicit time-dependence of $\theta(t)$ and $\varphi(t)$
is usually not indicated, but can be assumed.)
Since the vector $\vec\hat\omega$ of natural frequencies is not time-dependent,
the transformation is particular to this problem
and not really relevant to the model reduction technique discussed in this paper,
though it ensures the existence of a steady $\vec \theta$ state
for sufficiently high values of $\couplingK$.
This transformation is used to generate a new
dynamical system
\begin{equation}
    \label{eqn:trueRHS}
    \frac{\diff \theta_i}{\diff t}
    =
    \hat\omega_i
    -
    \frac{1}{\numNodes}
    \left[ 
    \sum_{j=1}^{\numNodes}
    \hat\omega_j \right]
    +
    \frac{\couplingK}{N}
    \left[
    \sum_{j=1}^{\numNodes}
    A_{i,j} \sin(\theta_j - \theta_i) \right]
    ,
\end{equation}
$\forall i\in[1, \numNodes-1]$.

The main idea involves mathematical ``technology transfer" from the field of \wordDefn{Uncertainty Quantification (UQ)}
\cite{Ghanem2003,Xiu2010}.
We assume that the long-time behavior of each unit in the assembly
is characterized by (is a function of) its {\em identity}--the value(s) of the heterogeneous parameter(s).
The problem can then be formulated
as being distributed in (heterogeneous) parameter space ($\vec p$-space)
in a manner analogous to spatiotemporal processes ``distributed" over physical space.
The state $\theta_i(t)$ for any unit $i$ with identity $\vec p_i$ can be approximated
in terms of appropriate basis functions
not in physical space, but rather in ``identity'' space: heterogeneous parameter space.
\begin{equation}
\label{eqn:gPCapproximant0}
\begin{array}{rccl}
\theta_i(t) & = & f(t;\vec p_i) \approx & \sum_{\scalarFuncIndex=1\ldots\numCofs}
    \alpha_{\scalarFuncIndex}(t)
    \basisFunc{\scalarFuncIndex} (\vec p_i).
\end{array}
\end{equation}
The basis functions  $\basisFunc{\scalarFuncIndex} (\vec p)$
are constructed as orthogonal polynomials in \secRef{polynomialGeneration}.
Because we are modeling cases in which the behavior of each unit
is assumed to be a smooth function of identity
(that is, nodes with similar identities are expected to behave similarly)
a relatively short truncation of such a series may well be accurate enough if the
right basis functions in $\vec p$ space are chosen.
In such a case, the number of ODEs to be solved
reduces from the number of units $\mathcal O(\numNodes)$, to the number of terms in the series $\mathcal O(\numCofs)$.
This approach, and its links to UQ modeling/computational developments (like the use of Smolyak grids for
nonintrusive collocation-based simulation) has been explored in \cite{Laing2012b}.
Yet these developments were only applicable for all-to-all coupled, \emph{intrinsically} heterogeneous assemblies of units.

The purpose of this paper is to generalize this approach by introducing a simple, yet nontrivial, extension.
Namely, we consider \emph{networks with non-trivial coupling structure}, i.e.~not all-to-all coupled.
In this work all connections have the same strength; further extension
to weighted connections is nontrivial, and the subject of current research.
Each unit is now also characterized, beyond its ODE parameter values,
by its \emph{connectivity} in the network---the nature of its coupling with other units in the network, which
in turn is quantified by 
features such as
the unit's {\em degree} 
(its count of undirected connections).
Different nodes have different connectivity features (imposed by the network structure)--we
can therefore think of connectivity as {\em a type of heterogeneity} of our building block units:
\wordDefn{structural} heterogeneity rather than \wordDefn{intrinsic} heterogeneity.

It should be noted that the steady state of similar dynamical systems have been predicted analytically \cite{Ichinomiya2004,Restrepo2005}.
However, \cite{Ichinomiya2004} considers the $N\rightarrow\infty$ limit, whereas we deal with finite $N$.
Additionally, \cite{Restrepo2005} requires that all the phases be known exactly
to use a self-consistency argument.
We are trying to obtain a reduced description,
so necessarily we do not know (nor want to know) all the phases of all oscillators,
but rather to approximate them.

In the system \eqnRef{kuramoto} to which we apply our coarse-graining strategy,
we will see that
the only connectivity feature that appreciably affects unit dynamics is
the unit degree
$\hat\degree_i=
\left(
    \sum_{j=1}^\numNodes A_{ij}
\right)
\in[0, N]$.
That is, though we have previously shown 
that intrinsically similar nodes in this system \cite{Moon2006}
(and others \cite{Moon2014})
have similar dynamics with all-to-all coupling,
we show here that,
with a nontrivial coupling topology network,
nodes which additionally have the same (similar) degrees
also have similar dynamics (possibly after a short initial transient).
The degree of a node can be treated
as another heterogeneous node parameter,
whose probability distribution is the network degree distribution.
We argue that the same approach which uses the distribution of an intrinsic heterogeneity,
and led to the reduction of all-to-all unit assemblies in \cite{Moon2006}
can be naturally extended to include a distribution over {\em structural heterogeneity}
that leads to reduction of unit assemblies coupled in networks.

We demonstrate this in the simplest nontrivial representative setting we can put together:
a set of coupled phase oscillators, characterized by heterogeneous frequencies $\hat\omega_i$
sampled from a prescribed distribution (here a truncated Gaussian)---%
but now \emph{not} all-to-all coupled.
Instead, the coupling $\mat{A}$ is in the form of a complex network,
generated by a Chung-Lu process similar to that described in \cite{Laing2012a,Chung2002}.
The ideas presented here work also for general networks,
if the degrees of the nodes are large enough;
the Chung-Lu network is used as a convenient example.
Likewise, the particular degree distribution tested here
(shown in \figRef{degreeHistogram})
is not itself important,
and networks with other degree distributions,
such as power-law,
or even nonmonotonic degree distributions,
can also be used.
Here, we first generate a weight sequence $w_i$, as
\begin{equation}
    \label{eqn:chungLuWeights}
    w_i = N  p  (1 - q (i-1) / N)^r\quad,\quad i=1,2,\ldots,N
\end{equation}
with parameters $p=\chungLuPVal$, $q=\chungLuQVal$, and $r=\chungLuRVal$.
With $\vec w$, we generate the connection probabilities $\mat P$ via
\begin{equation}
    \label{eqn:chungLuProbabilities}
    P_{ij} = P_{ji} = \min\left( \frac{w_i w_j}{\sum_k w_k}, 1 \right).
\end{equation}
We concretely generate an adjacency matrix $\mat A$ from these probabilities
by inverse transform sampling only above the diagonal of $\mat A$, and copying
to the bottom triangle, to ensure that the network is undirected,
with no self-loops.

Instead of following the behavior of each individual oscillator,
we exploit the observation that similar oscillators have
similar behavior and can be tracked together.
For all to all coupling, ``similar oscillators" is taken to imply similar natural frequencies,
and we write the oscillator state as a function of natural frequency \cite{Moon2006} and time.
However,
when a non-trivially structured coupling exists,
``similar oscillators" implies not only intrinsic similarity
but also {\em structural similarity}.
In addition to the intrinsic explanatory parameter of the natural frequencies,
the degree of each node appears to be an explanatory parameter
which suffices  (in our simple model) to capture the influence of the coupling structure
on the behavior of each oscillator;
yet other features such as in-degree or local clustering coefficient
are also worth considering.
If two oscillators have similar $\hat\degree$ values and similar $\hat\omega$ values, then their time-dependent behavior
is observed here to be similar, possibly after a short transient.
Finding the relationship between oscillator characteristics (intrinsic and structural)
and oscillator states generates a \wordDefn{coarse-grained} description,
whereby the system state can be encoded in fewer independent variables.

We illustrate a number of coarse-grained modeling tasks facilitated by this reduction: accelerated simulation
(via coarse projective integration), accelerated fixed point computation, continuation, and coarse-grained stability
analysis (via time-stepper based coarse Newton-Krylov GMRES \cite{Kelley1995,Kelley2003}
and Arnoldi algorithms \cite{Saad2011}%
).
This creates a natural link between the reduction approach we present here,
and our so-called Equation-Free framework for complex systems modeling
\cite{Kevrekidis2003,Kevrekidis2004}.

In the end, what makes it all possible is the fundamental assumption about how heterogeneity (intrinsic as well as structural)
affects the solution: ``nearby" parameter values and ``nearby" connectivities imply ``nearby" dynamics.
This is not always the case for {\em any} network, and so testing that this assumption holds
must be performed on a case-by-case basis.
For our networks, such a check is demonstrated in \figRef{unfitSurface}.
However, whether such a parameterization is possible
is linked to the question of whether frequency-synchronization emerges,
a subject for which an extensive literature exists \cite{Dorfler2014}.

The remainder of the paper is organized as follows.
First, we describe our illustrative example, a network of heterogeneous phase oscillators.
We then give the form of our low-dimensional representation of the system state.
We use our ability to transform back and forth between the two state representations
to perform several computational tasks in the coarse-grained space,
including the solution of initial value problems,
the computation of fixed points, their stabilities and bifurcations.
Appendices include an analysis of the validity of using higher-order coarse-grained integration schemes.

\section{An illustrative example of heterogeneous coupled oscillator networks}
\label{sec:exampleProblemSetup}

\begin{figure}[ht]
\centering
\begin{minipage}[b]{.45\linewidth}
    \centering
    \includegraphics[width=\textwidth]{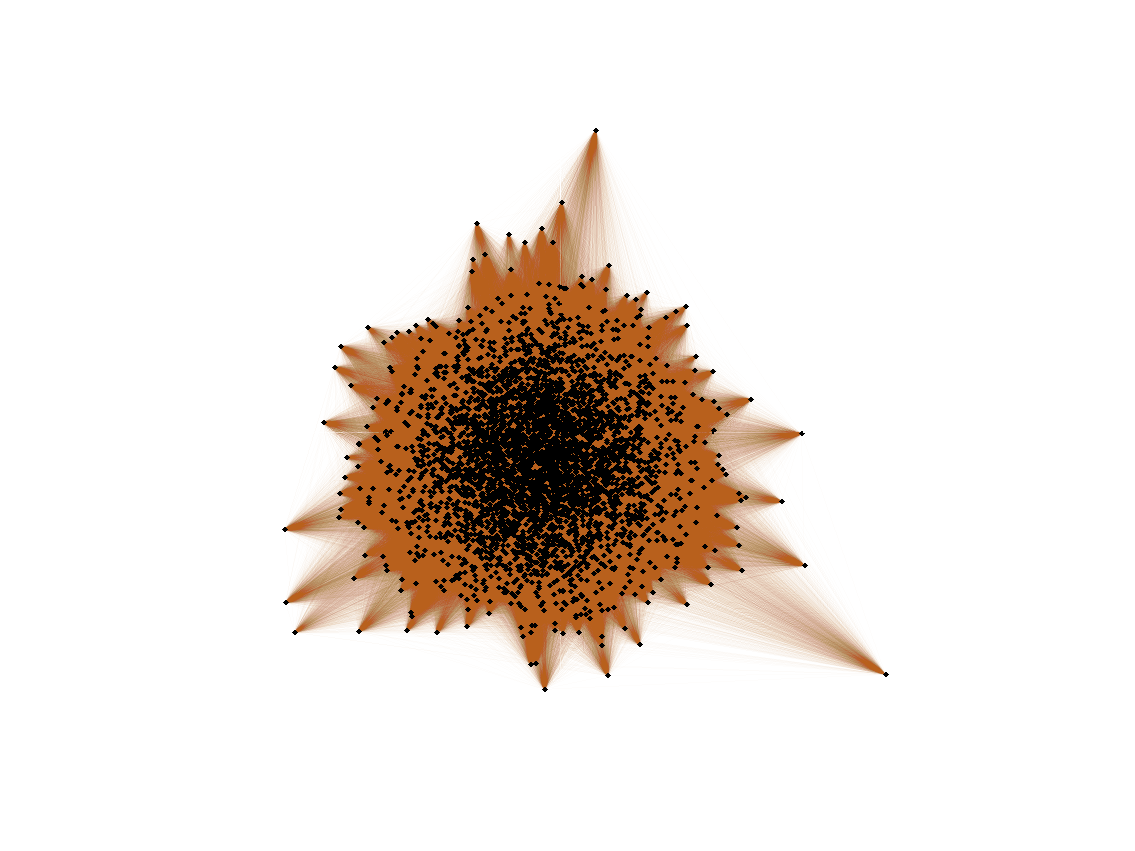}\subcaption{}\label{fig:graph}
\end{minipage}
\begin{minipage}[b]{.45\linewidth}
    \centering
    \includegraphics[width=\textwidth]{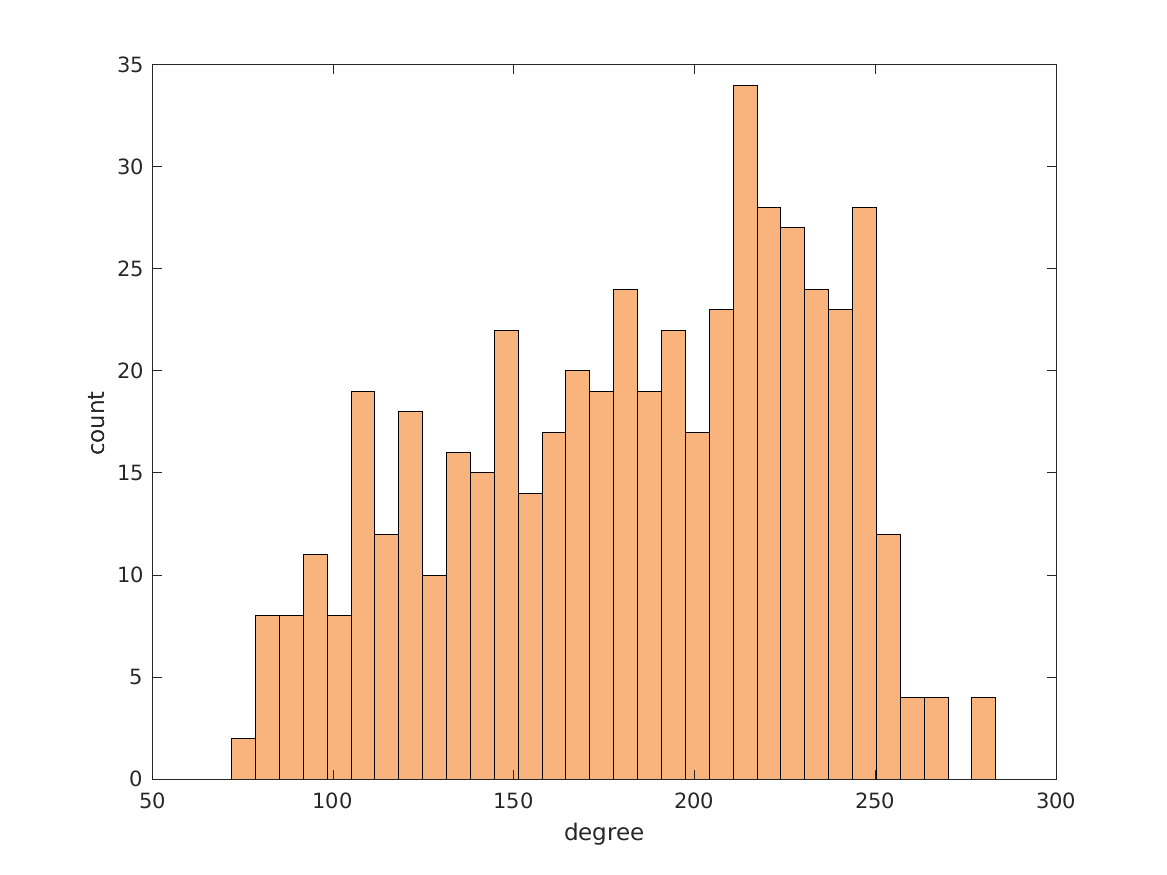}\subcaption{}\label{fig:degHist}
\end{minipage}
\caption{
    \textbf{Visualization of a Chung-Lu network} \cite{Chung2002} with $N=\graphPlotNval$ nodes,
    constructed using parameters $p=\chungLuPVal$, $q=\chungLuQVal$, and $r=\chungLuRVal$.
    In \ref{fig:graph}, the network is plotted with MATLAB's 2D spectral projection-based layout.
    In \ref{fig:degHist}, the degree histogram is shown.
}
\label{fig:degreeHistogram}
\end{figure}

\begin{figure}[ht]
\centering
\begin{minipage}[b]{.32\linewidth}
    \centering
    \includegraphics[width=\textwidth]{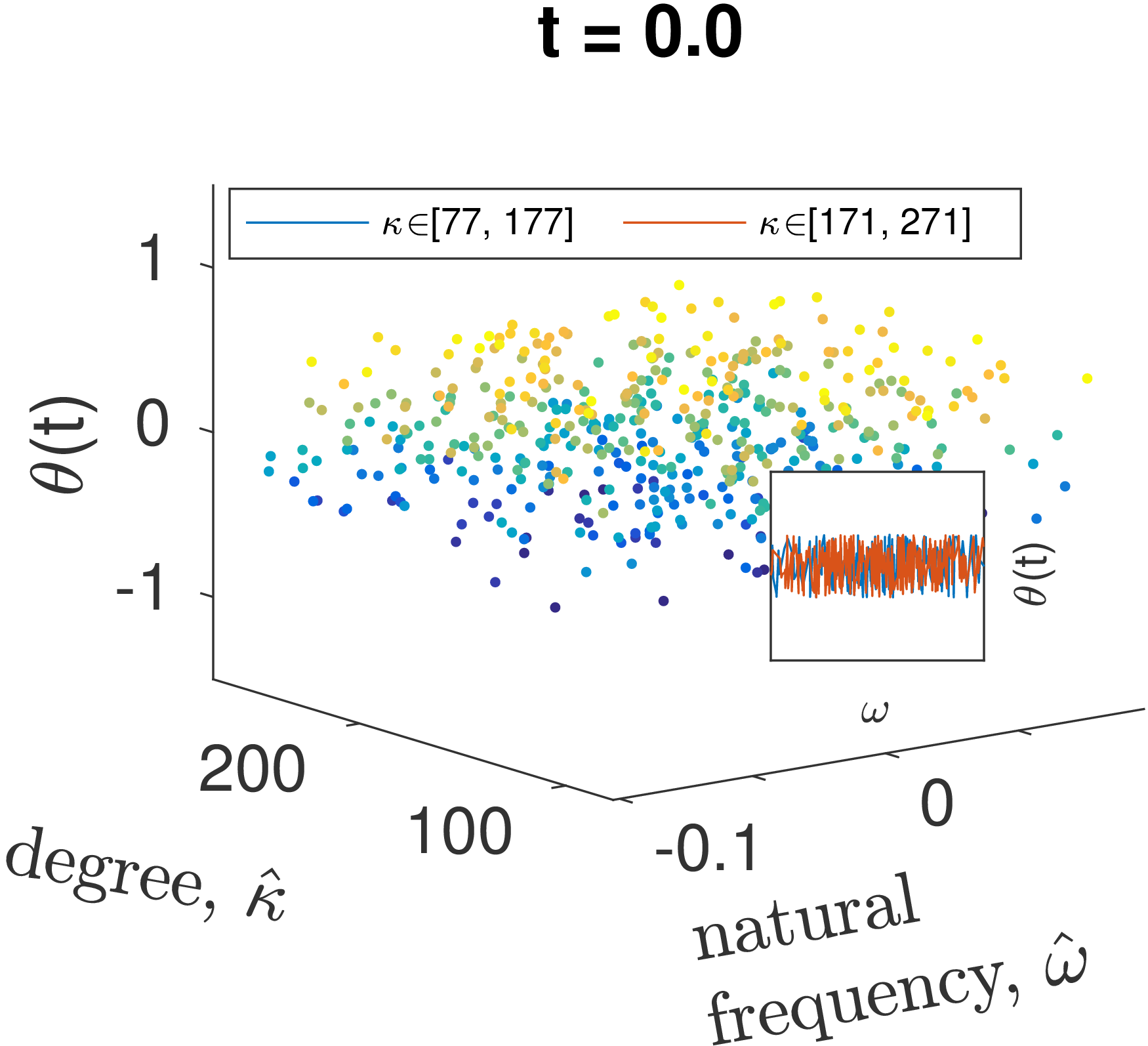}\subcaption{}\label{fig:plainScatter1}
\end{minipage}
\begin{minipage}[b]{.32\linewidth}
    \centering
    \includegraphics[width=\textwidth]{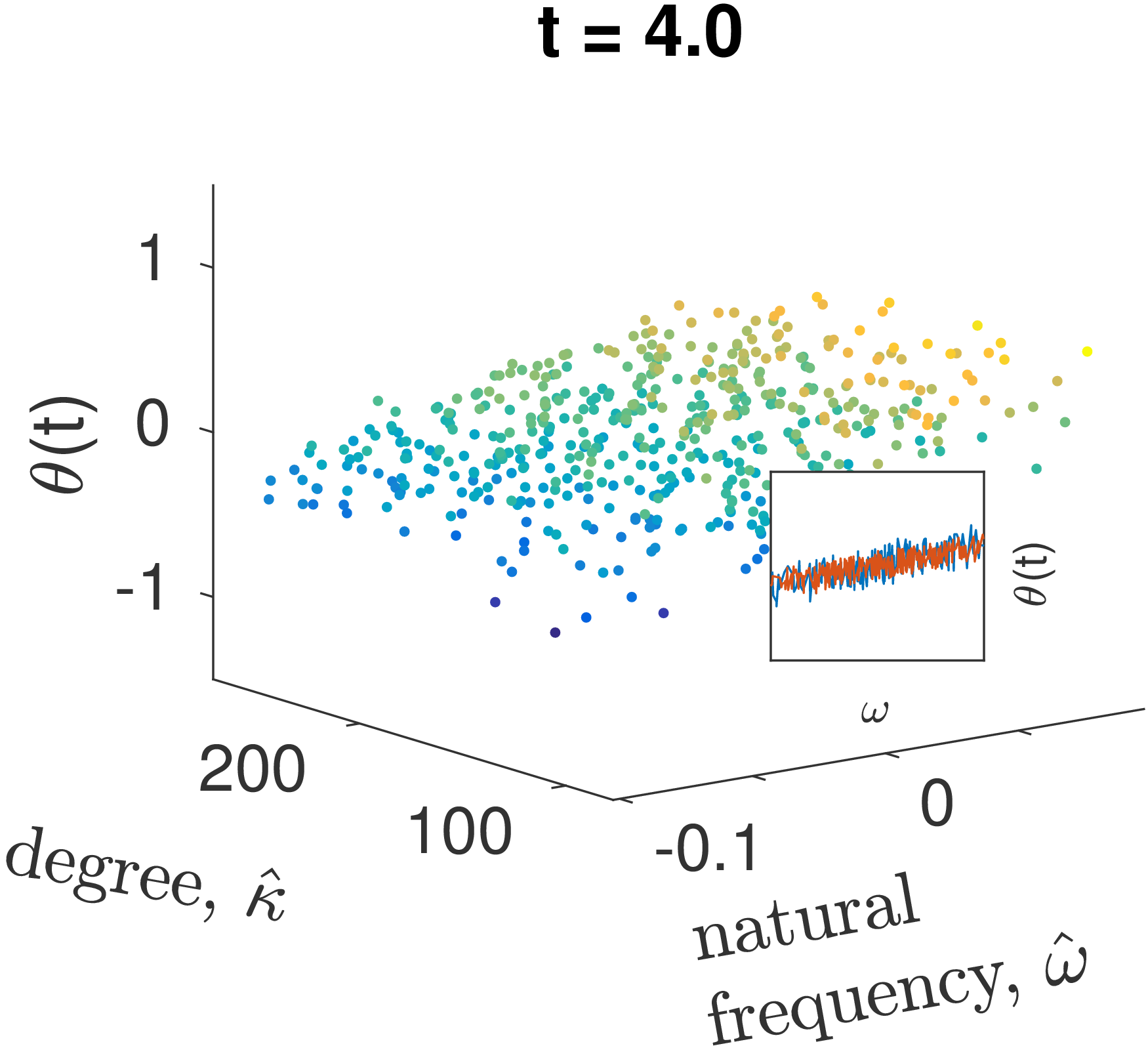}\subcaption{}\label{fig:plainScatter2}
\end{minipage}
\begin{minipage}[b]{.32\linewidth}
    \centering
    \includegraphics[width=\textwidth]{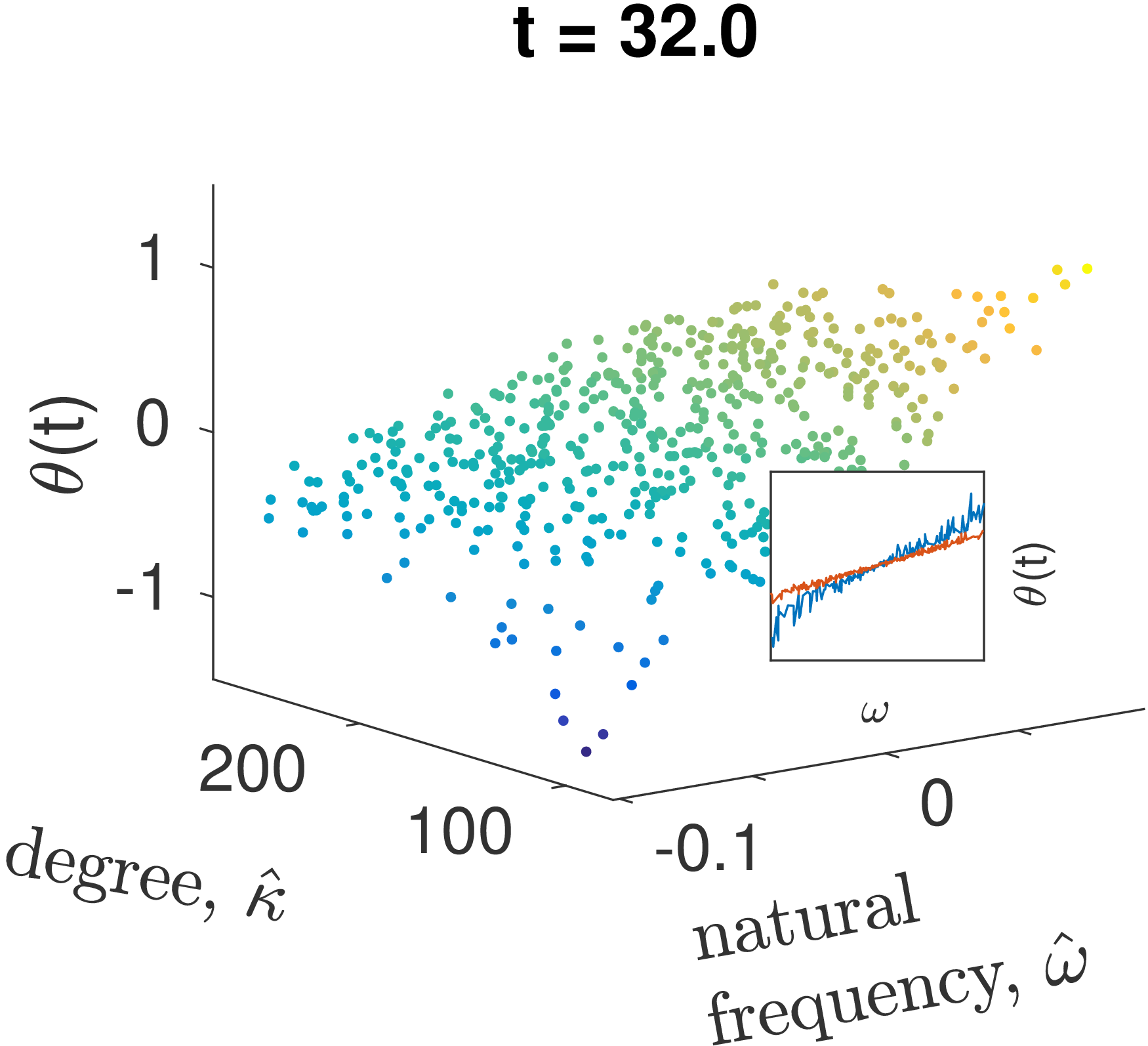}\subcaption{}\label{fig:plainScatter3}
\end{minipage}
\caption{
    \textbf{Oscillator states (phases) quickly slave to oscillator ``identities".}
    The network oscillator states
    are initialized as a cloud in $(\hat\omega,\,\hat\degree,\,\theta(t))$ space,
    and are clearly seen to quickly rearrange onto a 2-D surface.
    Points are colored by $\theta$ value.
    $N=\graphPlotNval$ and $\couplingK=\plainScatterKVal$ were used.
    Inset plots show slices at high and low $\hat\degree$ values, including all oscillators in two bands of width $\degreeSliceRadiusVal$.
}
\label{fig:unfitSurface}
\end{figure}

Our illustrative example is a network of coupled Kuramoto oscillators with
heterogeneous natural frequencies $\hat \omega_i$,
coupled in a stochastically generated network (here, a Chung-Lu network \cite{Chung2002}
with parameters $p=\chungLuPVal$, $q=\chungLuQVal$, and $r=\chungLuRVal$,
an example instance of which is shown in \figRef{degreeHistogram}).
This type of model system was used
in some previous reduction studies \cite{Moon2006,Rajendran2011}.
The number of oscillators in the network is also a parameter we will vary;
our base case is $N=\baseNVal$.
A basic premise, which is corroborated by \figRef{newtonResidual},
is that the network is large enough (the number $N$ of nodes
is large enough) for the single realization to be representative of the expectation
over all consistent network realizations.
The \wordDefn{fine} dynamics are governed
by the system of coupled ordinary differential equations (ODEs) \eqnRef{kuramoto},
where the natural frequencies $\hat\omega_i$ and the node degrees (numbers of neighbors)
$\hat\degree_i$
are heterogeneous across the oscillators constituting the network.
We remind the reader of our assumption that,
of all structural node features that may affect the dynamics,
it will be the node degree that matters here--so that the identity of node $i$
is sufficiently described by its intrinsic parameter $\hat\omega_i$
and its structural parameter $\hat\degree_i$.
This assumption is supported first by \figRef{unfitSurface},
and later, as we will see, more quantitatively by \figRef{newtonResidual}.

\setcounter{subfigure}{0}

We further define rescaled versions of the two heterogeneous parameters,
$x_i=(\hat x_i - \text{mean}(\{\hat x_j\}) / \text{stddev}(\{\hat x_j\})$ for
$x=\omega, \degree$
and
$i,j\in 1,\ldots,\numNodes-1$.
These two transformations do not affect the fine dynamics
of \eqnRef{kuramoto} or \eqnRef{trueRHS};
only the numerics of the implementation of the restriction $R$ to a coarse-grained state representation,
to be developed below.
Without axis markings, \figRef{unfitSurface}, for instance, would look the same
whether $\omega\times\degree$ or $\hat\omega\times\hat\degree$ were used for plotting.

The emergent functional dependence of the $\theta_i$ on the
$\omega_i$ (the intrinsic heterogeneity only) was discussed in the all-to-all coupling
context in \cite{Moon2005,Moon2006}.
There, we used a one-dimensional polynomial chaos expansion (PCE)
to describe the reduced problem
for $A_{ij}=1\,\forall i,j\in[1,\numNodes],\,i\neq j$,
so that all nodes
have degree $\hat\degree_i=\numNodes-1$.
In this paper, we again expect the oscillator states
to quickly become smooth (and time-dependent!)  functions of their identities,
but node $i$'s identity now includes \emph{both} $\hat\omega_i$ and $\hat\degree_i$.
For  $\couplingK$ sufficiently large so that a steady state of \eqnRef{kuramoto} exists,
we indeed observe that the states of randomly initialized oscillators quickly approach an
apparently smooth surface in $\omega \times \degree$ space
(see \figRef{unfitSurface})
suggesting that a low-order
series truncation of the type described in \eqnRef{gPCapproximant}
may constitute a good description.
This motivates the use of a functional fit of
the coefficients (the few $\alpha_{\scalarFuncIndex}(t)$ in \eqnRef{gPCapproximant})
to the data (the many $\theta_i(t)$)
as a coarse representation.

In previous work \cite{Rajendran2011} we have used a projection onto the eigenvectors of the discrete Laplacian
on the graph to describe the dependence of oscillator state on structural heterogeneity,
while using a one-term/linear fit
to account for dependence on intrinsic frequency.
\setcounter{subfigure}{0}

\section{Low-dimensional representation}

\subsection{Polynomial chaos}
\label{sec:polyChaos}
Given our observation that oscillator behavior quickly becomes
a function of oscillator {\em identity}, we want to describe the long-term
dynamics of the oscillator phase angles as a smooth function $\theta=\theta(t;\omega,\degree)$.
The phase angle of the $i$-th oscillator is then
given by $\theta_i(t) \equiv \theta(t;\omega_i,\degree_i)$.
Since our two heterogeneities
(the intrinsic and the structural) are here independent,
the basis functions are a tensor product of two independent polynomial bases 
\begin{equation}
 \label{eqn:basisFuncIndependent}
    \basisFunc{\multiIndex}(\omega, \degree)=\basisFuncSubscript{\multiIndexA}{\omega}(\omega) \basisFuncSubscript{\multiIndexB}{\degree}(\degree).
\end{equation}
This is a special case; the formulation will still in principle be applicable for parameters
with correlated joint probability distributions if one constructs an
appropriate set of basis functions \cite{Navarro2014,Deck2015}.

We now express the network dynamics in the form of a series expansion
in a truncation of this tensor product basis as
\begin{equation}
\label{eqn:gPCapproximant}
\begin{array}{rcl}
 \theta(t) & \approx & \sum_{\scalarFuncIndex=1}^\numCofs
    \alpha_{\scalarFuncIndex}(t) \basisFunc{\scalarFuncIndex} (\omega, \degree)
    \equiv \sum_{\scalarFuncIndex=1}^\numCofs
        \alpha_{\scalarFuncIndex}(t)
        \omegaOneDFunc^{ (\multiIndexA)}(\omega)
        \degreeOneDFunc^{(\multiIndexB)}(\degree)
    \\
    \indexSet&=&\{\multiIndex_{\scalarFuncIndex} = (\multiIndexA, \multiIndexB)\}
    :
    0 \le \multiIndexA, \multiIndexB \in\mathbb Z, \quad \multiIndexA + \multiIndexB \le p_\mathrm{max}
    \},
    \\
    \numCofs &=&||\indexSet|| = (1 + p_\text{max}) (2 + p_\text{max}) / 2
\end{array}
\end{equation}
where the $\alpha_{\scalarFuncIndex}(t)$ are time-dependent coefficients,
$\omegaOneDFunc^{(\multiIndexA)}(\omega)$
are basis functions arising from the intrinsic heterogeneity dependence
and $\degreeOneDFunc^{(\multiIndexB)}(\degree)$
are basis functions arising from the structural heterogeneity dependence.
Within the truncation of the set of functions $\indexSet$ included in the basis,
the ordering of the basis can be chosen arbitrarily, and so we substitute the
vector index $\multiIndex_{\scalarFuncIndex} = (\multiIndexA, \multiIndexB)$
with a scalar index $1 \le k \le \numCofs$.

The analogy with UQ now manifests itself in our
choice of the two independent basis sets: each one of them is chosen to be a polynomial chaos
basis in the corresponding heterogeneous (in analogy to random) parameter.
Each set of polynomials is orthogonal with respect to the probability density of the
corresponding heterogeneous parameter,
and the joint heterogeneity probability
density is just the product of the two unidimensional, independent heterogeneity probability
densities, so that \eqnRef{basisFuncIndependent} is satisfied,
as shown in \appendixRef{appendixProductOrthogonality}.

\newcommand{\pmax}{p_\mathrm{max}}
Note that in \eqnRef{gPCapproximant}, we specify that $\multiIndexA + \multiIndexB \le \pmax$.
This allows us to say that our two-dimensional polynomials are of {\em total degree}
$\le \pmax$.
An alternative truncation rule would be
to require that $\multiIndexA \le \pmax$ and $\multiIndexB \le \pmax$
(%
    or even to place separate bounds on $\multiIndexA$ and $\multiIndexB$,
    allowing for some anisotropy in the details,
    a topic for separate investigation%
).
Both approaches can be found in the literature.

This allows, per \eqnRef{gPCapproximant}, an approximation of the behavior as a time-dependent two-dimensional surface
in one intrinsic dimension (here, the (normalized) natural frequencies $\omega$),
and one structural dimension (here, the (normalized) node degrees $\degree$).
We repeat 
that this tensor product basis,
limited to those polynomials of total order less than some desired maximum,
is a truncated orthogonal basis for the 2D space
weighted by probability densities that are products of two marginal distributions.
Some standard distributions and their
corresponding families of orthogonal polynomials \cite{Xiu2010}
are given in Table \ref{tab:orthoPolys}.
\begin{table}[htbp]
\centering
\begin{tabular}{rlrl}
\multicolumn{2}{c}{\textit{Discrete}}
&
\multicolumn{2}{c}{\textit{Continuous}}
\\ \hline
\textbf{Distribution} & \textbf{Polynomials} & \textbf{Distribution} & \textbf{Polynomials}
\\ \hline
Binomial            & Kravchuk  & Gaussian  & Hermite \\
Poisson             & Charlier  & Gamma     & Laguerre \\
Negative binomial   & Meixner   & Beta      & Jacobi  \\
Hypergeometric      & Hahn      & Uniform   & Legendre
\\ \hline
\end{tabular}
\caption{Frequently encountered probability distributions
and the corresponding weighted orthogonal polynomial families.}
\label{tab:orthoPolys}
\end{table}

In classical Galerkin methods
an inner product is taken between the governing evolution equations and each basis function,
producing ODEs for the dynamics of the expansion coefficients 
(the $\alpha_{\scalarFuncIndex}(t)$ in \eqnRef{gPCapproximant}) by exploiting orthogonality.
A similar approach could be taken here
(through analytical computation of the inner product integrals if possible,
else through numerical quadrature).
Instead, we do not directly calculate
the temporal rates-of-change of the expansion coefficients,
but infer features of the coefficient dynamics
from brief bursts of simulation of the dynamics of the full system
(see \secRef{coarseTasks}).
This equation-free approach relies on our ability to go back and forth between
fine descriptions of the system state (the $\theta_i$ values),
and coarse ones (the $\alpha_{\scalarFuncIndex}$ values).
This is analogous to a non-intrusive (black-box, input-output) approach
to using polynomial chaos in UQ.

\subsection{Equation-free numerics}
\label{sec:polynomialGeneration}

The choice of the polynomial basis sets follows the selection of the appropriate
heterogeneity distribution.
For several frequently encountered distributions, the bases have been tabulated
(e.g. Table \ref{tab:orthoPolys}) from original generalized polynomial chaos
references (see e.g. \cite{Xiu2010}).
If not already available in such tables, one can construct the basis polynomials
e.g. through  Gram-Schmidt orthogonalization
with inner products in the space weighted by the distribution function.
A couple of nontrivial considerations arising in our case are that (a) our structural heterogeneity
parameter (the node degree) takes integer values, and so the degree distribution is discrete; and
(b) often we may encounter problems for which the heterogeneity distribution is not explicitly known,
but has to be estimated from specific system realizations (so, from large enough samples).
For the case of \emph{explicitly unknown} but \emph{sampled} distributions
(whether discrete or continuous)
we have used here the moments of the sampling of the heterogeneity parameters
for our particular network realization to extract the corresponding polynomials
(using SVD-based pseudo-inverses) \cite{Oladyshkin2012}.

\begin{figure}[ht]
\centering
\includegraphics[width=0.5\textwidth]{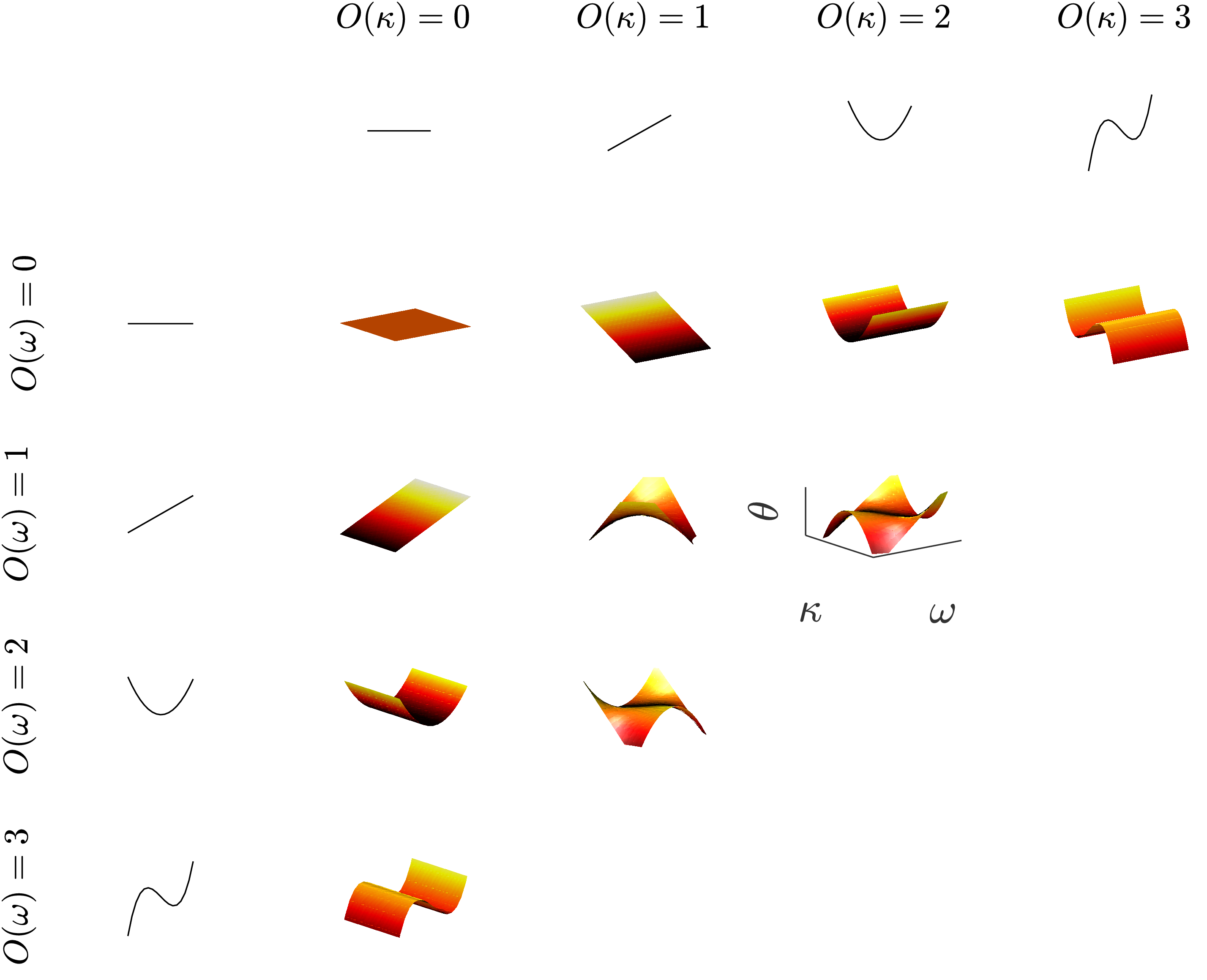}
\caption[Orthogonal $2D$ Polynomials]{
    \textbf{2D polynomials} (orthogonal with respect to the $\omega,\degree$ density 
    described in the text) are generated
    for a maximum total polynomial order of $p_\text{max}=3$.
}
\label{fig:2dPolyGrid}
\end{figure}

Results of the moment-based polynomial generation method,
which we used to generate all 1D polynomials used in this paper,
are shown in \figRef{2dPolyGrid}.
Here, the marginal samplings of degrees and natural frequencies
were used separately to generate 1D polynomials via moments,
and then a 2D basis was defined from the tensor product of these 1D bases,
with the restriction $\mathcal O(\omega) + \mathcal O(\degree) \le p_\mathrm{max}$
placed on the total polynomial degree of the 2D basis functions used.

For the expansion in \eqnRef{gPCapproximant},
we used a set of 2D basis functions.
\setcounter{subfigure}{0}
Once the relevant polynomials have been constructed, the appropriately defined inner product
also allows us (whether for continuous or for discrete distributions,
explicitly known or not)
to find the coefficients $\alpha_{\scalarFuncIndex}$ in \eqnRef{gPCapproximant}
for a given observation $\theta(t)$ of the system states
for a particular system realization (sampling of the distribution).
This can be accomplished directly (via numerical approximations of the
relevant inner products), using the orthogonality of the $\basisFunc{\scalarFuncIndex}$,
\begin{equation}
    \label{eqn:galerkinCoeffs}
    \alpha_{\scalarFuncIndex} = \frac{
        \intD{} f(\vec x) \basisFunc{\scalarFuncIndex}(\vec x) \diffMeasureVec
    }{
        \intD{} (\basisFunc{\scalarFuncIndex}(\vec x))^2 \diffMeasureVec
    },
\end{equation}
where we use the Lebesgue integral $\intD{} g(\vec x) \diffMeasureVec=\mathbb E[g(\vec x)]$.
For our problem, with one continuous and one discrete variable,
this can be written concretely as
$\intD{} g(\vec x) \diffMeasureVec
=
\int \sum_{\degree=0}^{\numNodes} g(\degree, \omega)
\rho_\degree(\degree)
\rho_\omega(\omega)
\diff \omega
$,
$(\omega, \degree)\in D$,
where $\rho_\degree$ is the (discrete) probability mass function for the degrees,
and $\rho_\omega$ is the (continuous) probability distribution function for the natural frequencies.
In separate work \cite{Choi2016}, we examine the computation of this integral
for the case when the problem can be recast as a PDE,
and so the coupling sum in \eqnRef{kuramoto} and \eqnRef{trueRHS}
can also be written as an Lebesgue integral.
There, we consider standard Monte Carlo integration
in addition to Gaussian quadrature and a repurposing of anchored ANOVA.
These latter methods have the benefit of allowing integrals to be computed
using only a few key virtual oscillators,
with anchored ANOVA having the additional benefit of decreased scaling sensitivity
to the number of random dimensions (2 in this paper).
However, these benefits require that the original model be recast as PDEs
continuous in both time and the random dimensions.

Here, we take the alternate approach of finding the $\alpha_{\scalarFuncIndex}$
indirectly through least squares fitting,
minimizing the squared residual norm $\sigma$
with respect to the coefficients $\alpha_{\scalarFuncIndex}$
(here using a QR algorithm).
\begin{equation}
\label{eqn:errMinimizedByProjection}
\begin{array}{rcl}
\sigma &=& || f(\vec x) - \sum_{{\scalarFuncIndex}=1}^{\numCofs}
                    a_{\scalarFuncIndex} \basisFunc{\scalarFuncIndex}(\vec x)
                    ||_2^2 \\
&\approx& \hat \sigma = \sum_{i=1}^\numSamps w(\vec x_i)
\left(
f(\vec x_i) - \sum_{{\scalarFuncIndex}=1}^{\numCofs}
                    \alpha_{\scalarFuncIndex} \basisFunc{\scalarFuncIndex}(\vec x_i)
                    \right) ^ 2,
\end{array}
\end{equation}
where $\lim_{\numSamps\rightarrow\infty} \hat \sigma = \sigma$
according to the law of large numbers,
and the weights $w(\vec x_i)$ are still to be decided.

For a (large) finite sample $\vec x_i$, $i=1,\,\ldots,\,\numSamps$,
if we take the partial derivative $\partial \hat \sigma / \partial \alpha_{\scalarFuncIndex}$,
and use the fact that
$
    \sum_{i=1}^\numSamps w(\vec x_i)\basisFunc{\scalarFuncIndexAlt}(\vec x_i) \basisFunc{\scalarFuncIndex}(\vec x_i) = 0
$
if ${\scalarFuncIndexAlt} \neq {\scalarFuncIndex}$ (orthogonality) to remove some terms, then we find that
\begin{equation}
    \alpha_{\scalarFuncIndex} = \frac{
        \sum_{i=1}^\numSamps w(\vec x_i) f(\vec x_i) \basisFunc{\scalarFuncIndex}(\vec x_i)
    }{
        \sum_{i=1}^\numSamps w(\vec x_i) ( \basisFunc{\scalarFuncIndex}(\vec x_i) )^2
    }.
\end{equation}
So, as long as we accept that
\begin{equation}
    \label{eqn:gMean}
    \sum_{i=1}^\numSamps w(\vec x_i) g(\vec x_i)
\end{equation}
is a good approximation to
\begin{equation}
    \label{eqn:gExpectation}
    \intD{} g(\vec x) \diffMeasureVec,
\end{equation}
we obtain the same formulas for the $\alpha_{\scalarFuncIndex}$.

Suppose $\vec x$ has a density $\rho(\vec x\in D)$, so \eqnRef{gExpectation} can be written as
\begin{equation}
    \label{eqn:gExpectationRho}
    \intD{} g(\vec x) \rho(\vec x) \diff \vec x.
\end{equation}
If the $\vec x_i$ are chosen randomly in accordance with $\rho(\vec x)$,
then \eqnRef{gMean}, where $w(\vec x_i)=1/\numSamps$,
is a good approximation to \eqnRef{gExpectationRho}.
This is just Monte Carlo integration, and the law of large numbers gives
\begin{equation}
    \label{eqn:lawOfLargeNumbers}
    \lim_{\numSamps\rightarrow\infty}
    \frac{1}{\numSamps}
    \sum_{i=1}^\numSamps g(\vec x_i)
    =
    \intD{} g(\vec x) \rho(\vec x) \diff \vec x.
\end{equation}

\section{Coarse Computational Modeling Tasks}
\label{sec:coarseTasks}

Beyond their conceptual simplification value, collective (coarse) variables
can be valuable in facilitating the computer-assisted study of
complex dynamical systems by accelerating tasks such as direct simulation,
continuation, stability and bifurcation analysis for different types of
solutions.
To accomplish this acceleration, the Equation-Free approach \cite{Kevrekidis2003,Kevrekidis2004}
is predicated on the ability to map between corresponding fine and coarse
descriptions of the same system.

This is accomplished through the definition of a \wordDefn{restriction operator}
$R : \mathbb R^{\numNodes-1} \rightarrow \mathbb R^\numCofs$
which maps from fine states $\vec \theta(t)$
to corresponding coarse states $\vec\alpha(t)$
by minimizing the residual $\sigma(\vec \alpha(t))$ from \eqnRef{errMinimizedByProjection}.
We also need to define the counterpart of restriction: a \wordDefn{lifting operator}
$L : \mathbb R^\numCofs \rightarrow \mathbb R^{\numNodes-1}$
which maps $\vec\alpha$ vectors to $\vec\theta$ vectors
by setting the $\theta_i$ values equal to the right-hand-side of
the approximant in \eqnRef{gPCapproximant}
evaluated at the corresponding $(\omega_i, \degree_i)$.

One more important thing to note before proceeding to demonstrating the approach is that
the lifting operator is, in general, a one-to-many relation; there are many
fine realizations of the process that are mapped to the same
coarse representation---coarse-graining (e.g. averaging) loses information.
If the problem can be usefully coarse-grained, \emph{any} of these consistent fine
realizations, or the average of several of them, can be used practically in the definition
of the coarse time-stepper below; we may think of the coarse-timestepper as the
{\em expected value} over all such consistent realizations.
In singularly perturbed multiscale problems one can clearly see how the memory
of the details of the lifting are quickly forgotten, suggesting that any
consistent fine realization is ``good enough" to estimate this expectation \cite{Gear2002,Kevrekidis2009}.

The $L$ and $R$ operators combine to define a
coarse timestepper $\Phi_{{\inTstep}, C}$, in
\begin{equation}
 \label{eqn:coarseTimestepper}
 \begin{array}{rcl}
 \Phi_{{\inTstep},C} &:& \mathbb R^\numCofs \rightarrow \mathbb R^\numCofs
 \\
 \Phi_{{\inTstep},F} &:& \mathbb R^{\numNodes-1} \rightarrow \mathbb R^{\numNodes-1}
 \\
 \Phi_{{\inTstep},F}[\vec\theta(t)]
 &=& \vec\theta(t+{\inTstep})
 \\
 &=&
 \int_{s=t}^{s=t+{\inTstep}}
 \frac{\diff \vec \theta(s)}{\diff t}
 \diff s
 \\
 \Phi_{{\inTstep},C}[\vec\alpha(t)]
 &=&
 \vec \alpha(t + {\inTstep})
 \\
 &\equiv&
 \left(
 R
 \circ
 \Phi_{{\inTstep},F}
\circ
 L
 \right)
 [\vec\alpha(t)]
 \end{array}
\end{equation}
This is the timestepper for the (unavailable) coarse-grained dynamical system, approximated
through observing the results of short bursts of appropriately
initialized fine-grained simulations.
A single evaluation of this coarse time-stepper, by itself, does not provide any computational
savings; it is the way we design, and process the results of, several such
coarse time-steps that leads to computational benefits.
Using traditional numerical analysis codes (initial value solvers, fixed point solvers)
as templates for wrapper codes around the coarse timestepper, tasks like
accelerated simulation, coarse-grained stability and bifurcation analysis,
optimization, and controller design, can be performed.
This wrapper technology is described in detail (and fruitfully used to explore
model coarse-graining across disciplines) in a series of publications \cite{Theodoropoulos2000,Kevrekidis2010}.
What is important here is not the established wrapper algorithms technology;
it is the selection of coarse observables, leading to the appropriate definition
of the coarse time-stepper, that makes the entire program feasible and useful.

\subsection{Coarse Initial Value Problems}
\label{sec:IVP}

We can use the coarse timestepper to accelerate
the computation of dynamic trajectories of the system,
through Coarse Projective Integration (CPI) \cite{WilliamGear2003,Lee2007}.
Given a coarse initial condition $\vec\alpha(t=0)$ we lift to
a consistent fine scale state $L[\vec\alpha(t=0)]$ and use it to
initialize a fine scale numerical integrator.
We run for a short time $\inTstep$ (the inner step) and
we record the final coarse state by restricting the corresponding fine state, $R[\vec\theta(t + \inTstep)]$.
We use these two coarse states to estimate the coarse time derivative, which
we then use in the forward Euler formula to project forward in time the coarse state
for a (large, coarse) time step $\outTstep$ (the outer step).
This constitutes the simplest coarse projective forward Euler integration scheme:
\begin{equation}
    \label{eqn:CPIeuler}
        \vec \alpha(t+\outTstep)
        =
        \vec \alpha(t)
        +
        \outTstep
        \frac{
            \vec\Phi_{\inTstep,C}[\vec\alpha(t)]
            -
            \vec\alpha(t)
        }{\inTstep}
\end{equation}

A slightly more sophisticated approach
would also take two points separated by an inner step size $\inTstep$
to approximate the rate-of-change of $\vec \alpha$,
but only after first performing a healing integration
in the fine equations
\cite{Gear2002}.
If the lifted representation of the state $\vec \alpha(t)$,
as projected from the previous timestep $\vec \alpha(t-\outTstep)$,
is slightly off the hypothetical slow manifold in the fine space,
this short healing trajectory dampens the fast components
which are not captured in the coarse representation
\cite{Vandekerckhove2011,Antonios2012,Gear2004}.

It is important to note here that in all our CPI computations we used
a single network (with a single $\omega$ vector)  to generate the polynomial 
basis functions {\em and} to lift to at every projective step.
This can be thought of as a ``single instance" CPI; one may also consider
CPI for the expected behavior over all networks that share the same 
degree as well as $\omega$ distributions -- in which case one should lift
to many consistent network realizations and average over them.
This issue will be examined more closely below.

\begin{figure}[ht]
\centering
\includegraphics[width=0.5\textwidth]{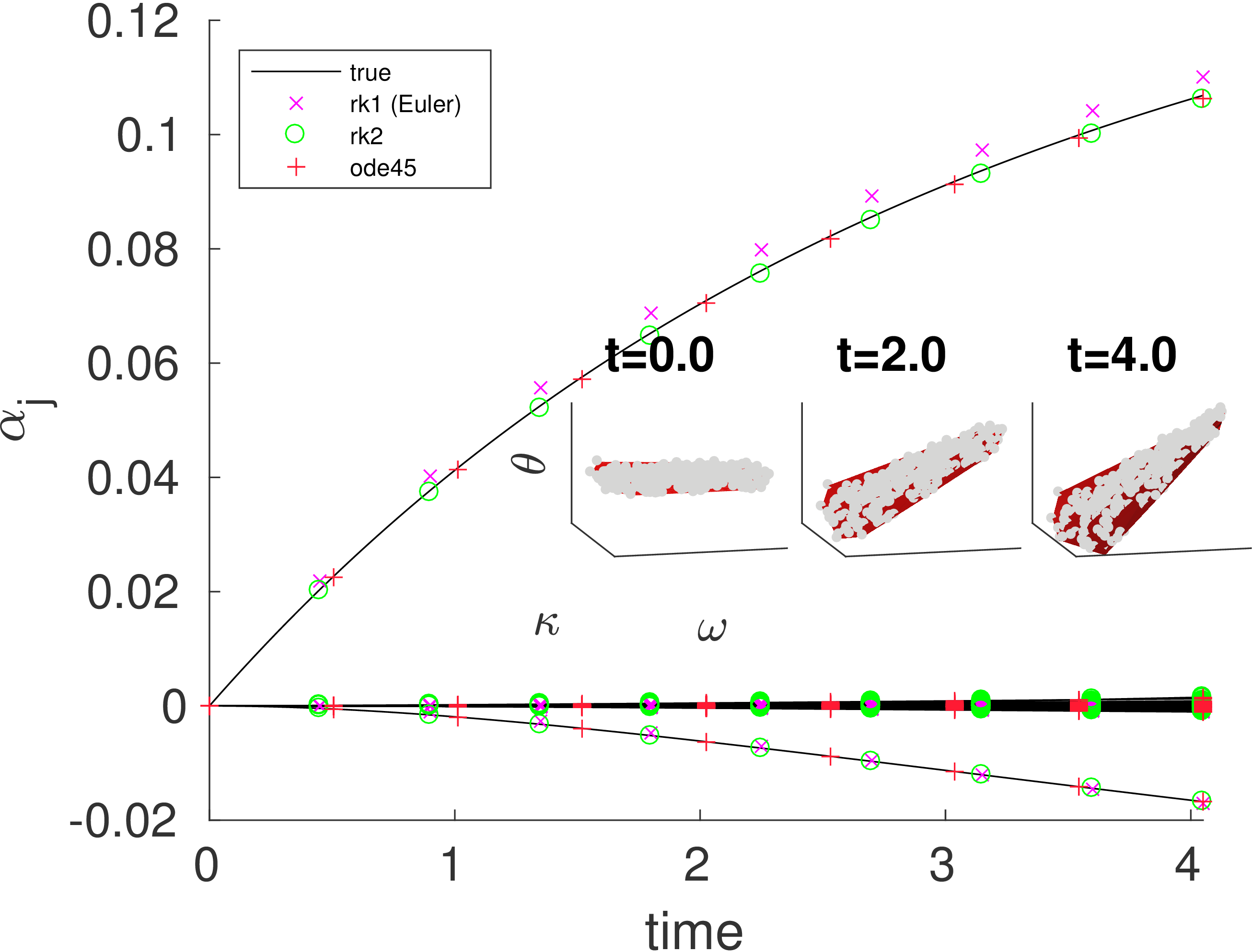}
\caption{
    \textbf{Coarse projective integration} shows smooth evolution of the first few leading
    PCE coefficients $\alpha_{j}$,
    with some corresponding fine states visible in the 3D insets.
    Black curves (in the main figure) and dense grey scatters (in the insets) were obtained by full fine integration
    with the same initial $\theta$ conditions using MATLAB's \texttt{ode23}.
    Colored points (in the main figure) and red surfaces (in the insets) were obtained via CPI,
    using several different integrators.
    At each coarse step, $\frac{\diff \alpha_{\scalarFuncIndex}}{\diff t}$ was estimated
    $\forall {\scalarFuncIndex}=1,\ldots,\numCofs$
    (where $\numCofs = \numCofsVal$)
    by drawing $\numCofs$ chords through the restrictions of the last two points
    in a brief burst of fine integration of ${\inTstep} = \flowTimeVal$ time units.
    At the times indicated,
    we make inset plots with red surfaces corresponding to the lifted CPI state
    and grey scatters corresponding to the closest (in time) state in the true trajectory.
    These should  be compared to \figRef{unfitSurface}.
    We performed the same task for several outer integrators:
    two explicit Runge-Kutta integration schemes,
    and a coarse wrapper around the built-in MATLAB integrator \texttt{ode45}
    are compared to the restrictions
    of points in the fine trajectory starting from the same lifted initial condition.
    For the two explicit Runge-Kutta integrators,
    an outer step of
    $\outTstep=\KuramotoCPItStepVal$
    was used.
    For \texttt{ode45}, an absolute
    tolerance of $10^{\KuramotoCPIatollogTenVal}$
    and a relative 
    tolerance of $10^{\KuramotoCPIrtollogTenVal}$
    were used.
    $N=\cpiNval$, $\couplingK=\defaultKVal$, and $\numCofs=\numCofsVal$ were used.
    The $\omega$ values were drawn from a truncated normal distribution
    supported on $[-\hiOmVal, \hiOmVal]$, with zero mean and standard deviation $\omStdVal$.
    }
\label{fig:cpiResults}
\end{figure}

\begin{figure}[ht]
\centering
\includegraphics[width=0.5\textwidth]{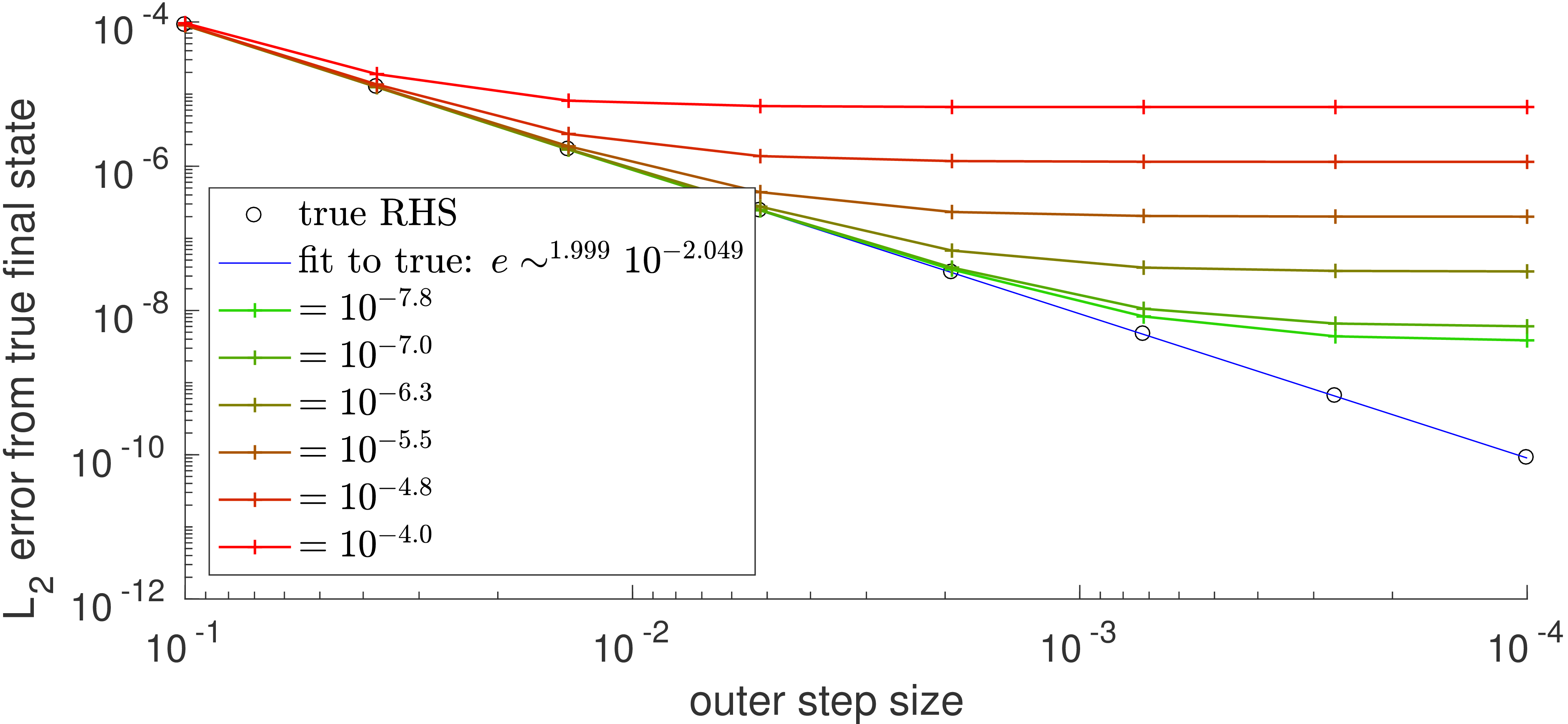}
\caption{
    \textbf{As the inner step size ${\inTstep}$ decreases,
    the error (compared to direct integration) of a projective (not coarse-projective) integration
    becomes bounded by the integrator's intrinsic (outer) step size ${\outTstep}$.}
    The true solution at $t=\dhdtDeltaTVal$ was found by integrating
    using MATLAB's \texttt{ode45} with an
    absolute tolerance of $10^{\dhdtatolLogTen}$ and a 
    relative tolerance of $10^{\dhdtrtolLogTen}$
    The series of black circles give the error at $t=\dhdtDeltaTVal$
    that results from using integration using the true RHS function \eqnRef{trueRHS}
    in an explicit second-order Runge-Kutta integration scheme
    of (outer) step size ${\outTstep}$.
    The colored curves use the same integrator and outer step size,
    but approximate the RHS function with the difference map
    $f_{\inTstep}(\theta(t)) = \theta(t) - \Phi_{{\inTstep},F}[\theta(t)]$,
    analogous to the coarse difference map of \eqnRef{coarseDifferenceMap}.
    Error was evaluated by taking the norm of the vector difference
    between the projective integration solution $\vec \theta(\dhdtDeltaTVal)$
    and the true solution.
    Compare this to \figRef{hTau2D} in the Appendix, in which a similar analysis is performed
    on a system of two ODEs modeling a single reversible reaction.
}
\label{fig:hTauPI}
\end{figure}

Results of applying the simpler scheme
to the coupled oscillator problem
with coarse variables obtained by a $2D$ PCE fit
are shown in \figRef{cpiResults}.
In general, the coarse time-stepper results can be used to estimate
the coarse right-hand-side function $\dot{\vec \alpha}$,
which is not available in closed form.
On demand numerical estimates of this right-hand-side through
short bursts of appropriately initialized fine simulation allow us to use other existing integrators,
such as MATLAB's \texttt{ode45},
to computationally approximate coarse trajectories (also shown in \figRef{cpiResults}).
\appendixRef{appendixRKConv2} contains a quick illustration of the useful properties of such projective
initial value solvers.
It is shown there that, under reasonable conditions, the order of a \emph{projective} integrator
templated on a two-step Runge-Kutta initial value solver (including the additional estimation step for the coarse time derivatives)
is the same as the order of the
\emph{actual} Runge-Kutta initial value solver.
\figRef{hTauPI}) confirms this for projective
integration of the fine equations for our model.
\setcounter{subfigure}{0}
\setcounter{subfigure}{0}

This approximation of the coarse right-hand-side function
can be used for other computational tasks
besides the computation of dynamic trajectories.

\subsection{Coarse Fixed Point Computation}

The coarse time-stepper can be used to define a coarse difference 
\begin{equation}
\label{eqn:coarseDifferenceMap}
\vec F_{\inTstep}[\vec\alpha(t)]
=
\vec \Phi_{{\inTstep},C}[\vec\alpha(t))]
-
\vec\alpha(t).
\end{equation}

Steady states of the fine timestepper are clearly zeroes
of this difference;
one expects that the zeroes of the coarse difference
correspond to coarse steady states of the original problem.
$F_{\inTstep}$ can therefore be used to find coarse steady states
involving only $\numCofs$ variables.
Iterative, matrix-free linear algebra lends itself to finding zeroes
of such a problem in the absence of explicit equations for the dynamics
of the coarse variables $\alpha_j$,
We used a Krylov-type matrix free technique (Newton-Krylov GMRES)
to converge to such coarse steady states.

\begin{figure}[ht]
\centering
\begin{minipage}[b]{.45\linewidth}
    \centering
    \includegraphics[width=\textwidth]{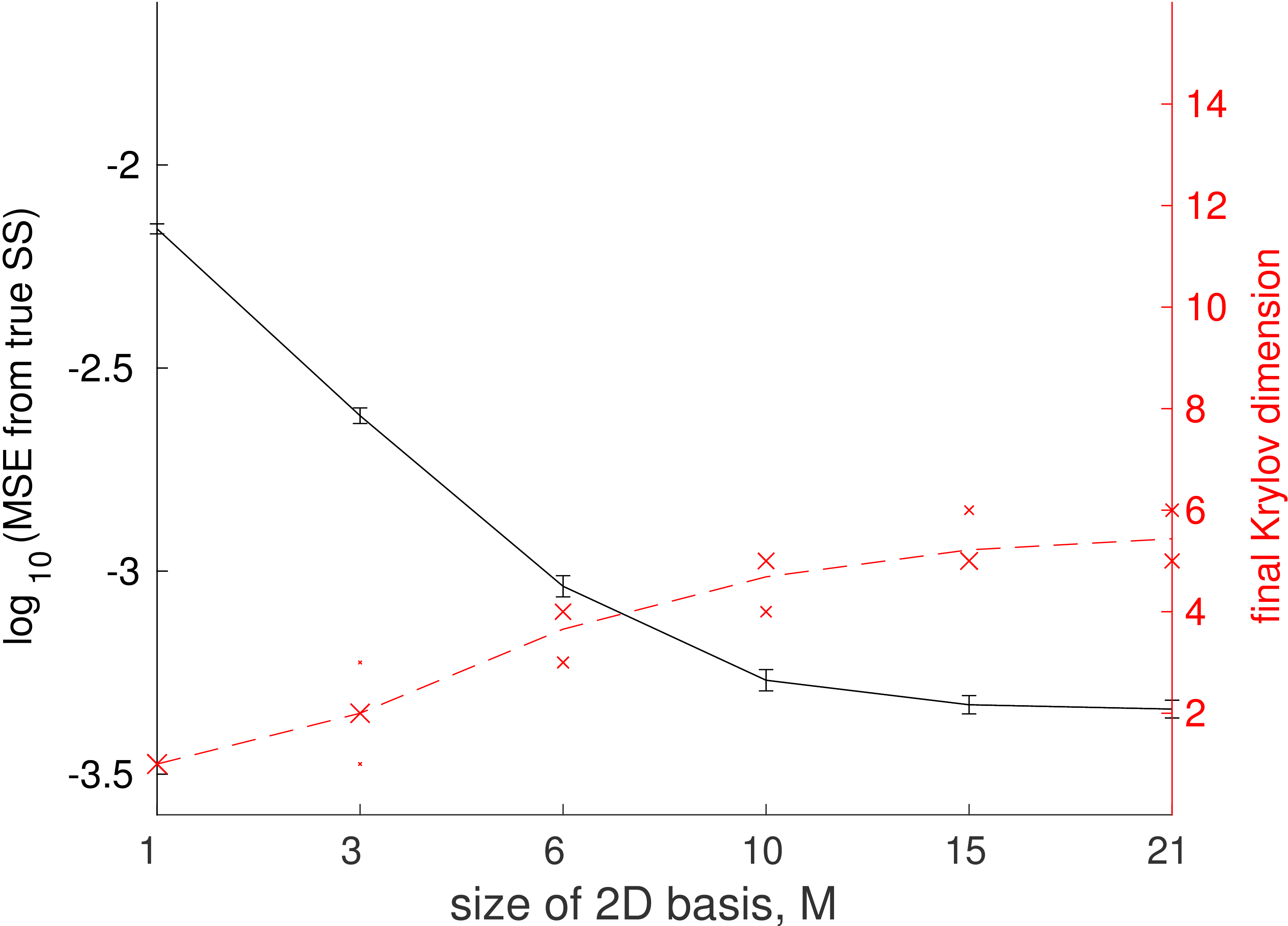}\subcaption{}\label{fig:newtonResidual}
\end{minipage}
\begin{minipage}[b]{.45\linewidth}
    \centering
    \includegraphics[width=\textwidth]{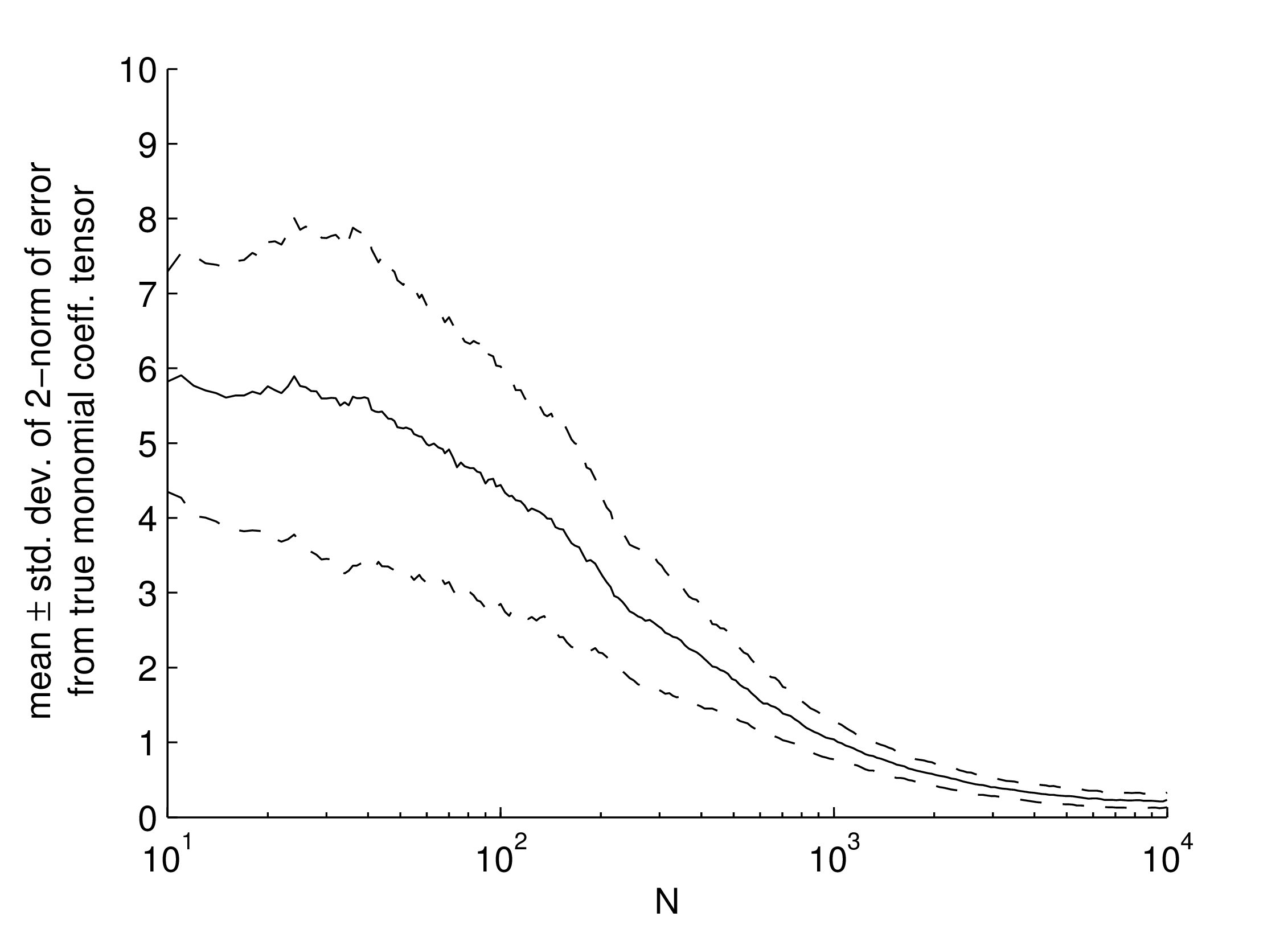}\subcaption{}\label{fig:polyErrorN}
\end{minipage}
\caption{
    \figRef{newtonResidual}
    \textbf{ Coarse fixed point computations, and their sensitivity to the size of the coarse basis used.}
    Computations were repeated for different samples of network and natural frequencies and
    the error bars are indicative of the resulting variation in the solution.
    Error was computed between lifted fixed points of \eqnRef{coarseDifferenceMap} (with ${\inTstep}=\flowTimeVal$),
    and steady states of \eqnRef{trueRHS}.
    The norm used was the mean squared error (MSE) across the $N-1$ nodes.
    \numReplicatesVal{} replicates were used per value of the independent variable,
    an absolute tolerance of $10^{\absTolPowVal}$ was used for the outer Newton solver,
    error bars are \errBarTypeVal,
    and $\couplingK=\defaultKVal$ was used.
    \figRef{polyErrorN}
    \textbf{Convergence of the orthogonal polynomials based on (increasingly larger) finite networks.}
    As $\numNodes$ increased, the polynomials generated via sample moments approached
    the $\numNodes=\bigBigNVal$ polynomials.
    Error was quantified in the 2-norm
    of the monomial coefficient tensor $\mat C$
    in $\basisFunc{\scalarFuncIndex}(\omega, \degree) =
    \left(
     \sum_{{\scalarFuncIndexAlt}=0}^{p_{\scalarFuncIndex,\omega}} C_{\omega,{\scalarFuncIndex},{\scalarFuncIndexAlt}} x^{{\scalarFuncIndexAlt}}
     \right)
     \left(
     \sum_{{\scalarFuncIndexAlt}=0}^{p_{\scalarFuncIndex,\degree}} C_{\degree,{\scalarFuncIndex},{\scalarFuncIndexAlt}} x^{{\scalarFuncIndexAlt}}
     \right)
     $,
    where the pair $\vec p_{\scalarFuncIndex}$ gives the orders of the two one-dimensional polynomials.
}
\label{fig:coarseSteadyStateConvergence}
\end{figure}

A Newton-Krylov iteration to find such a state is depicted in
\figRef{coarseSteadyStateConvergence}.
In Newton-Krylov GMRES (Generalized Minimal RESidual),
the inner linear problem of an outer (nonlinear) Newton-type solver
is solved by GMRES,
in which the solution to $\mat B \vec x = \vec b$
is assembled in a space
derived from the $n^{th}$ Krylov subspace $\{\mat B^j \vec r_0 \}_{j=0,\ldots,n-1}$,
where $\vec r_0$ is the residual of the initial iterate
(and $\mat B$, which is not computed, is the Jacobian of \eqnRef{coarseDifferenceMap}).

Here we work again with $\numNodes=\baseNVal$ node networks.
The basis polynomials are computed from {\em a single} network realization (a large,
$\bigBigNVal$ node Chung-Lu network); because the support of the degree distributions for
a $\numNodes=\bigBigNVal$ and for a $\numNodes=\baseNVal$ network are not the same,
the degrees and frequencies are normalized as described in \secRef{exampleProblemSetup}.
But now we construct \numReplicatesVal{} realizations of networks consistent with the chosen degree
distribution, and perform our fixed point computation for each one of them.
We do {\em not} regenerate polynomials for each of these realizations; we observe them
on the ``large sample" polynomials; this is justified in \figRef{polyErrorN},
where we recover the Probabilist's Hermite polynomials as $\numNodes$ increases,
by generating with sample moments.
The effect of regenerating polynomials within a larger computation
is considered further in \cite{Rajendran2012,Rajendran2016}.

We report the sensitivity of the results to the basis size $\numCofs$ in \figRef{newtonResidual},
where the error bars are indicative of the variation across our $\numReplicatesVal$ network samples.
We can also see in \figRef{newtonResidual} that the dimension of the Krylov subspace
(at the final Newton iteration before convergence) initially closely follows
the size of the 2D basis we use (it uses "all of $\numCofs$") but later on plateaus.
\setcounter{subfigure}{0}

\subsection{Coarse stability computations: Eigenvalues and eigenvectors}

(Coarse) eigenvalues of the Jacobian of the (coarse) difference map \eqnRef{coarseDifferenceMap}
upon convergence to its (coarse) fixed points
can be used to establish the stability of said fixed points and help
determine the nature of their potential (coarse) bifurcations.
These eigenvalues $\coarseEig_i$
are related by
\begin{equation}
\label{eqn:eigM2L}
\trueEig_i \approx {\coarseApproxdEig}_i = \ln(\coarseEig_i + 1) / {\inTstep}
\end{equation}
to the corresponding eigenvalues ${\coarseApproxdEig}_i$ of the Jacobian of the (unavailable)
coarse differential evolution equations; in turn, these should coincide with the
leading eigenvalues $\trueEig_i$ of the actual problem (the leading eigenvalues of
the  \emph{fine} differential equations).
For $\numCofs>\eigMethodThreshVal$ coarse eigenvalues $\mu_i$ were obtained through the Jacobian-free
implicitly restarted Arnoldi Method (IRAM)
\cite{Kelley1995,Lehoucq1998}
applied to the coarse difference operator \eqnRef{coarseDifferenceMap}.
For $\numCofs\le\eigMethodThreshVal$,
a forward finite-difference Jacobian with a fixed step size of $\fdJacDiffVal$
was computed and the eigenpairs calculated directly
with the QZ algorithm implemented in MATLAB's \texttt{eig}.

\begin{figure}[ht]
\centering
\begin{minipage}[b]{.45\linewidth}
    \centering
    \includegraphics[width=\textwidth]{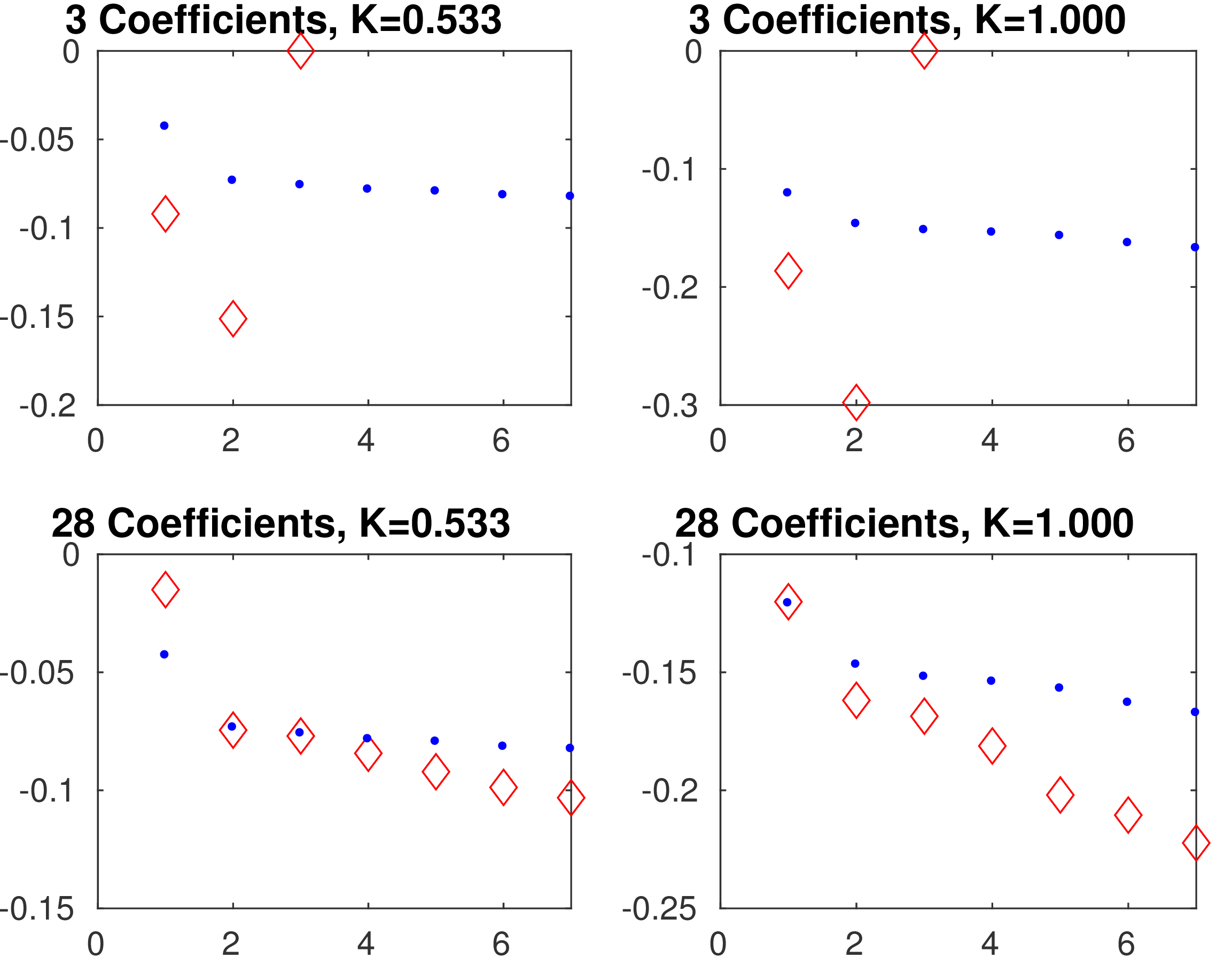}\subcaption{}\label{fig:evals}
\end{minipage}
\begin{minipage}[b]{.45\linewidth}
    \centering
    \includegraphics[width=\textwidth]{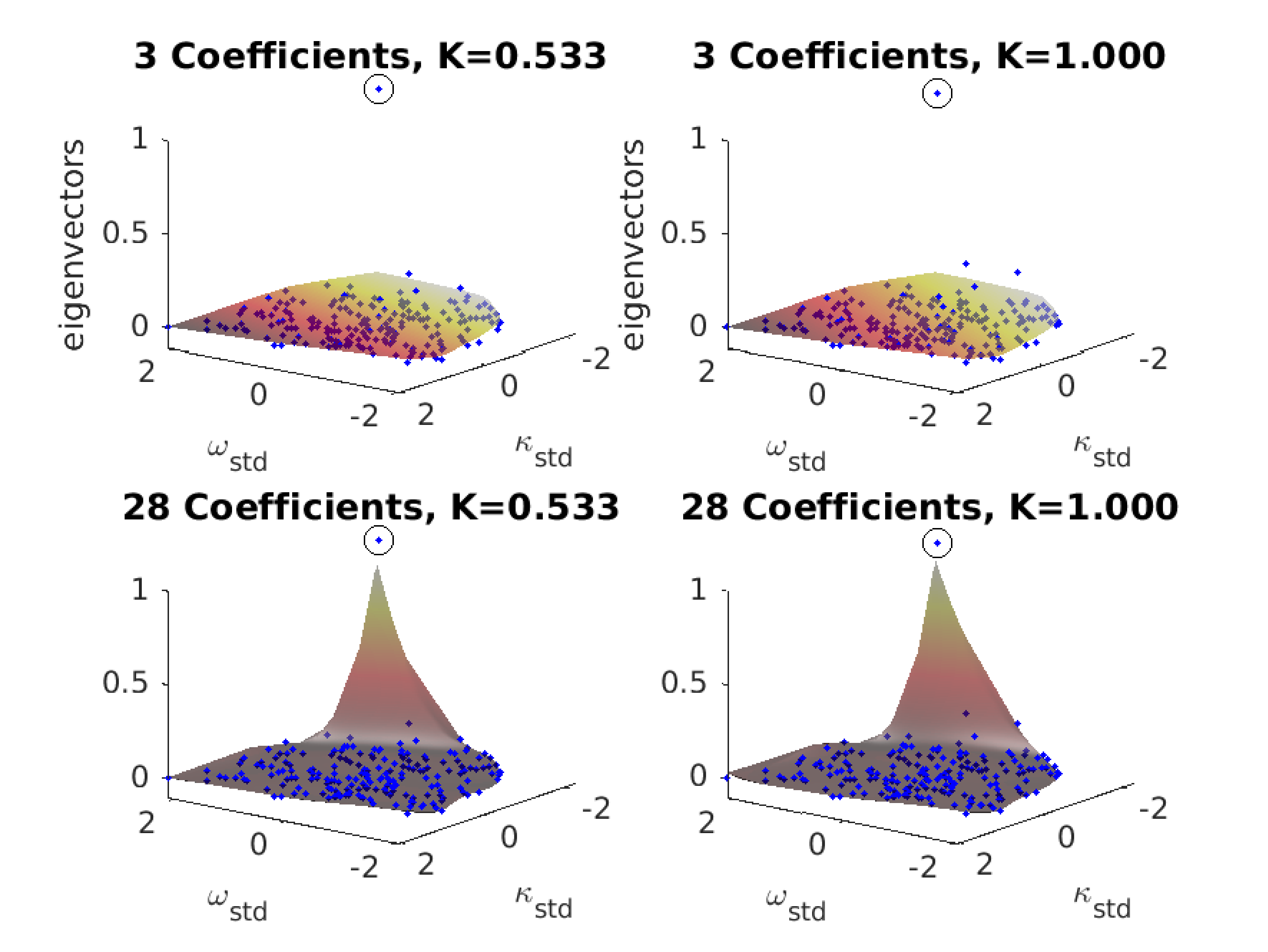}\subcaption{}\label{fig:evecs}
\end{minipage}
\caption{
    \textbf{
    Comparison of coarse and fine eigencomputations for
    different basis sizes close to (left column) and
    far away from (right column) the main SNIPER bifurcation (see text).
    }
    Eigenpairs obtained
    from \eqnRef{coarseDifferenceMap}
    (with ${\inTstep} = \flowTimeVal$,
    $N = \eigcompsNval$,
    and $\numCofs = \eigcompsMval$)
    are similar to those obtained from \eqnRef{trueRHS}.
    True eigenvalues and eigenfunctions (small blue points) were obtained using MATLAB's \texttt{eig}
    on an analytical Jacobian of \eqnRef{trueRHS}.
    Approximate eigenvalues (large red diamonds) and eigenvectors (smooth surfaces)
    were obtained via implicitly restarted Arnoldi iteration (IRAM),
    with the transformation \eqnRef{eigM2L}.
    converge to the the fine eigenvalues $\lambda_i$
    as $M$, the number of $\alpha_{\scalarFuncIndex}$ coefficients, rises.
    For larger values of $\couplingK$,
    the eigenvalues are all increasingly negative,
    though the ratio between the first and second eigenvalue (about $0.6$)
    does not change by much.
    \\
    In \figRef{evals}, the horizontal axis of each plot is an index across eigenvalues,
    while the vertical axis is eigenvalue.
    In \figRef{evecs}, the lifted view of the leading coarse eigenvector visually
    approaches the leading fine eigenvector.
    The coarse eigenvector was evaluated as a surface
    (in a manner similar to \eqnRef{gPCapproximant})
    at a fine grid of points within the convex hull of the sampled $(\omega, \degree)$ points.
    The ``eigensurface" corresponding to the slowest eigenvalue appears to approach an indicator function on the oscillator
    whose extreme $(\omega,\degree)$ pair makes it the most susceptible to ``desynchronization"
    with decreasing $K$.
    Eigenpairs were chosen to match the right (synchronized) inset plot in \figRef{bifurcation},
    and the point closest to the turning point
    along the branch of coarse fixed points in that figure.
}
\label{fig:eigeninformation}
\end{figure}

As the number of coarse variables is increased,
and therefore the quality of the coarse approximation improves,
one expects these coarse eigenvalue estimates to approach the leading eigenvalues of
the analytical fine Jacobian 
of \eqnRef{trueRHS}, located through any eigensolver.
In \figRef{evals}, we demonstrate
this convergence of the approximate eigenvalues $\coarseApproxdEig_i$ to 
the leading fine eigenvalues $\trueEig_i$
with increasing $\numCofs$.
\figRef{evecs} shows the corresponding
convergence of a coarse eigenvector (the one
corresponding to the smallest absolute value of $\coarseApproxdEig_i$)
to 
the fine eigenvector corresponding to smallest absolute value of $\trueEig_i$.
Note that lifting is necessary to make a comparison between $\theta$ and $\alpha$ eigenvectors.
On the other hand, the transformation \eqnRef{eigM2L} is necessary
not because of our coarse and fine spaces,
but because the $\coarseEig$ eigenvalues come from the Jacobian of a finite-time flow map
while the $\trueEig$ eigenvalues come from the Jacobian of a vector of infinitesimal differential equations.
\setcounter{subfigure}{0}

In performing computations involving finite differences, we used a value of
$\sqrt{\text{machine precision}}$ (according to \cite{Press1987}), which is approximately $10^{-7}$
for IEEE standard double-precision floating point variables in MATLAB.

\subsection{Coarse Continuation/Bifurcation Diagrams}

\begin{figure}[ht]
    \centering
    \includegraphics[width=0.5\textwidth]{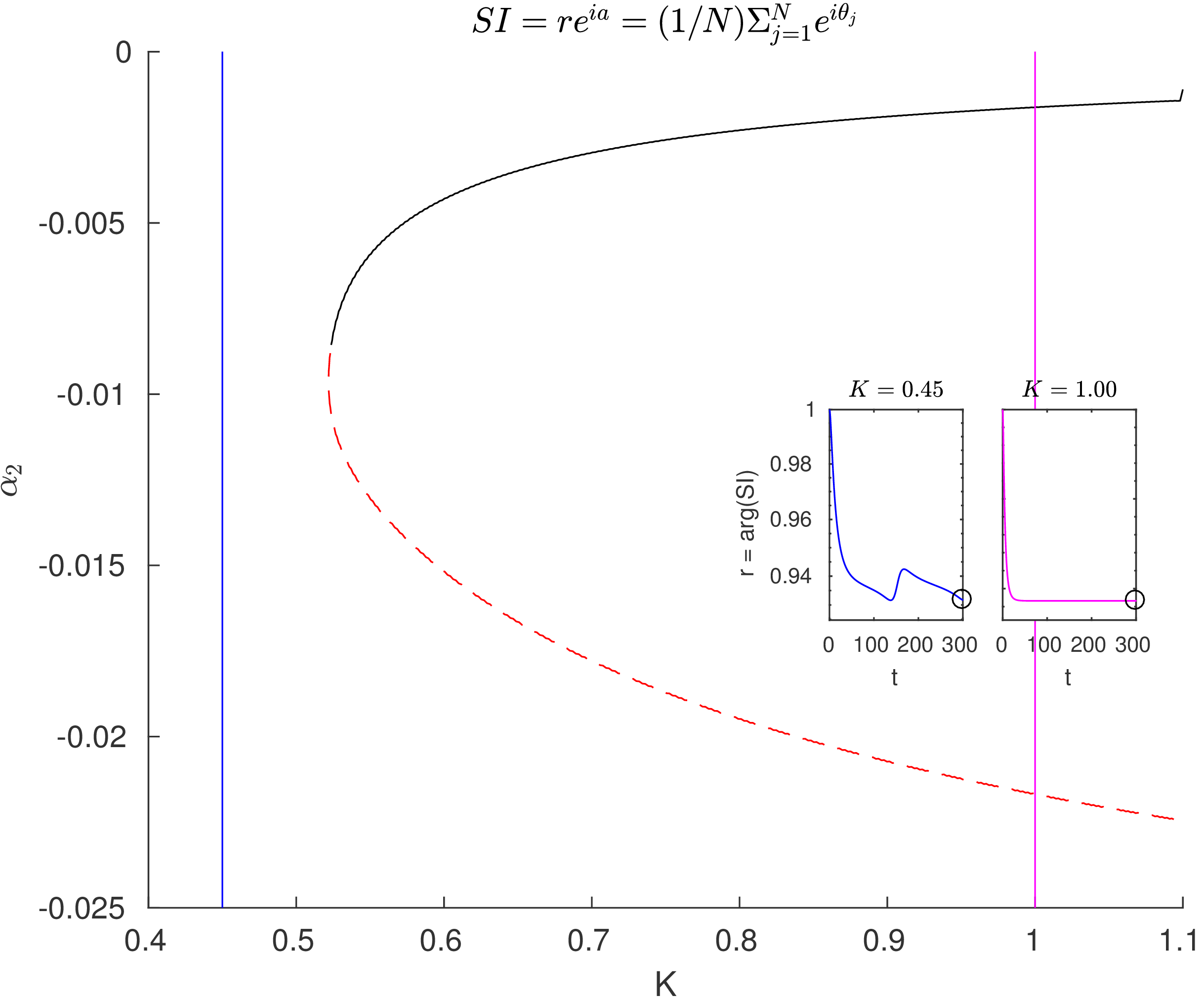}
    \caption{
    \textbf{Bifurcation Diagram, Coarse and Fine.}
    Coarse fixed-point solutions
    can be used to generate a bifurcation diagram in the parameter $\couplingK$,
    via pseudo-arclength continuation of fixed points of the coarse flow map
    in \eqnRef{coarseTimestepper}
    (with ${\inTstep}=\pseudoArclengthTauVal$,
    $N=\eigcompsNval$,
    and $\numCofs=\eigcompsMval$).
    At the point where the color of the curve changes from red to black,
    one eigenvalue passes through zero.
    This marks a change from a stable to an unstable branch.
    The two inset plots show representative trajectories
    of the real magnitude $r$ of the complex synchronization index
    $r e^{ i a } = \frac{1}{\numNodes} \sum_{j=1}^\numNodes e^{i \theta_j}$.
    This quantity can be thought as a vector pointing to the mean phase angle,
    whose length approaches $1$ as the oscillators approach perfect (phase-)synchronization.
    In the right half of the bifurcation diagram, trajectories approach stable steady states,
    where oscillators are completely (frequency-)synchronized.
    In the left half, one (or more) rogue oscillator(s) travel(s)
    around the phase ring alone, slowing briefly
    when passing through the cluster of synchronized oscillators.
    For both insets, the initial condition was $\theta_i=0\forall i$.
    }
    \label{fig:bifurcation}
\end{figure}

To build a coarse bifurcation diagram (See \figRef{bifurcation}),
we performed pseudo-arclength continuation \cite{Keller1977,Kelley2005} for the coarse
fixed points.
We computed branches of coarse solutions to
$\vec 0 = F_{\inTstep}(\vec \alpha; \couplingK)$ as the global parameter $\couplingK$ is varied.
To trace out these solution branches, steps were taken in (pseudo-)arclength along the branch
rather than in $\couplingK$.
This allows the continuation to extend naturally beyond turning points.
At some point along this continued branch of solutions,
one of the computed eigenvalues becomes positive.
At this point, the line color is changed to indicate that the new branch
comprises unstable solutions.
\setcounter{subfigure}{0}

Beyond these branches (to the left) we know that a limit cycle solution
arises:  a periodic orbit characterized by one free \wordDefn{rogue oscillator},
which performs full rotations and only momentarily slows down as it passes through 
the remaining
pack of clustered oscillators \cite{Moon2005}.
In dynamical systems terminology this is a ``SNIPER" (saddle-node infinite period) 
bifurcation \cite{Strogatz2001}.
The insets in \figRef{bifurcation} show transient dynamics
in terms of the \wordDefn{synchronization index} $\SI$ in
$\SI e^{i\avgPhase} = \frac{1}{\numNodes}\sum_{j=1}^\numNodes e^{i \theta_j}$
(the real magnitude of the complex Kuramoto order parameter; see e.g. \cite{Skardal2011}).
The presence of this rogue oscillator means that
the coarse representation of \eqnRef{gPCapproximant}
is not particularly accurate/informative to the left of $\couplingK_c$,
without explicitly including the rogue's value of $\theta$
in the set of coarse variables,
as was done, for example, in \cite{Moon2006}.

\section{Discussion}

In this paper we have demonstrated that a general network of coupled, intrinsically heterogeneous, oscillators
can be usefully described using a small number of collective dynamic variables. 
These variables are the
time-dependent coefficients of an expansion
of the complete state of the network in terms of a set of orthogonal polynomials. 
The polynomials are products of
univariate polynomials in the parameters describing the \emph{intrinsic heterogeneity} of a given oscillator, 
and a \emph{structural} heterogeneous property (here, the degree) 
indicative of the connectivity of the oscillators in the network.
Our results extend previous work which
only considered all-to-all coupled networks, in which the state of
an oscillator was a function of {\em only} its intrinsic heterogeneity \cite{Moon2006,Laing2016}.
Our expansion (and subsequent truncation of the expansion) in this form
is motivated by the large body of work in the field of uncertainty
quantification; the difference being that here we have heterogeneous parameters characterizing a single network,
rather than many realizations of a dynamical system, each with different (uncertain) parameters.
We anticipate that this new link between the two fields (network dynamics and UQ) may provide
many more fruitful opportunities for mathematical/computational technology transfer that 
can enhance our understanding and ability to usefully describe and analyze dynamics on complex networks.

Although we have only considered Kuramoto-type oscillators in a specific Chung-Lu network,
our methods do not rely on either the type of oscillator used or the specific network (as long as the mean degree is not small). 
Thus they should be widely applicable to many non-trivial networks of neurons which exhibit synchrony for some range of parameters

Using this reduced description of a network, we demonstrated a number of standard computational tasks using the
equation-free framework, in which differential equations describing the evolution of the expansion coefficients
are not explicitly derived, but rather estimated on-the-fly. 
Specifically, we demonstrated coarse projective integration,
the computation of coarse fixed points and their stability, 
as well as parametric analysis through continuation. 

The success of our method relied on the rapid development of correlations between the state of an oscillator
and its heterogeneous identifying parameters, in this case, its intrinsic frequency and its degree. 
A potential shortcoming of the method would arise when such a strong dependence does not develop--that is,
when ``similar" oscillators do not behave ``similarly"
(e.g. when the initial conditions, or something more than just the degree,
like the clustering coefficient of every node, matters).
This implies that additional ``heterogeneity dimensions" must be introduced,
in analogy to when, say, a two-dimensional flow loses stability and becomes three-dimensional.
One such case we have encountered \cite{Moon2014} is when
the oscillators in a network (an all-to-all network of Hodgkin-Huxley neurons) split in two subsets 
and in each subset a distinct relation of state to identity was established.
Knowing the identity of the oscillator was not enough, in that case, to characterize dynamics; one also needed to know in
which cluster the oscillator belonged. 
We could regard both our case as well as this other one as special cases where, for every oscillator identity,
there is a distribution of behaviors--a strongly peaked unimodal distribution in this paper, 
and a strongly peaked bimodal distribution in \cite{Moon2014}; for that matter, in our study of breakup in multiple 
communities/clusters, one obtained a multimodal distribution. 
We anticipate that, in the spirit of stochastic PDEs in physical space, our approach might be extended to
evolve state distributions in heterogeneity space (as opposed to state functions in heterogeneity space).

In other networks
it may be that the state of a node depends on more than just these two properties: networks with weighted
edges provide an obvious context in which this may occur.
We believe (and are actively pursuing this research direction) that the
approach introduced here can also be usefully extended to help in determining reduced descriptions
for such networks.

The tensor product basis used here relied on a lack of correlation across the heterogeneities.
As mentioned in \secRef{polyChaos}, this reliance can be overcome
by generating the full multidimensional basis all at once,
and our current work addresses this possibility.
This is likely to be the case in dynamical systems
in which it is is useful to retain multiple structural heterogeneities.
Degree is one of several structural parameters--others include 
a node's participation in motifs like
triangles (complete graphs on three nodes),
cherries (triangles with one edge removed)
or its local clustering coefficient.
This progression can be  continued
to higher-order statistics of the node connectivity
by noting that using the degree of each node as the representative structural heterogeneity
is equivalent to considering the per node counts of the two-node one-edge motif.
As more structural heterogeneities are considered,
it is reasonable to expect that these heterogeneities\
will not be statistically independent.

\section*{Author Contributions}
TB, IGK, CL and CWG planned the work, and TB and YW performed the computations.
All the authors contributed to the writing of the manuscript.

\section*{Funding}
This work was partially supported by the US National Science Foundation and
by NIH grant 1U01EB021956-01.

\section*{Acknowledgements}
IGK, TB and CL gratefully acknowledge the hospitality of the
Institute of Advanced Studies of the Technical University of M{\"u}nich at Garching
where IGK is a Hans Fischer Senior Fellow.
TB and IGK thank Thomas Thiem for useful comments/discussions.

\bibliographystyle{natbib}
\bibliography{main}

\newpage
\appendix

\section{Orthogonality of the tensor product basis for independent weightings}
\label{sec:appendixProductOrthogonality}

Here, we demonstrate that if we have polynomials in two variables that are orthogonal
with respect to a two-dimensional weight function
that is the product of two one-dimensional weight functions,
then the orthogonal polynomials
are the product of the respective one-dimensional orthogonal sets.

\subsection{Problem setting}
We suppose that we are given two one-dimensional orthogonal function sets
defined on (possibly infinite) intervals $I_x$ and $I_y$
based on the weight functions $\rho(x)$, $x\in I_x$,
and $\sigma(y)$, $y \in I_y$
and we want to construct a set of two-variable orthonormal polynomials,
$\basisFuncSymbol_{i,j} (x, y)$
of degrees $i$ in $x$ and $j$ in $y$
such that
\begin{equation}
    \label{eqn:innerIMJN}
    \langle \basisFuncSymbol_{i,j}, \basisFuncSymbol_{m,n} \rangle_{\rho\sigma} = \delta_{im} \delta_{jn}
\end{equation}
where
\begin{equation}
    \langle f, g\rangle_{\rho\sigma}
    =
    \iint f(x,y) g(x,y) \rho(x)\sigma(y)\diff x \diff y
\end{equation}
and
    $\{\phi_i(x)\}$ and $\{\theta_j(y)\}$
are the orthonormal sets of polynomials corresponding to the
one-dimensional systems based on the weights
$\rho(x)$ and $\sigma(y)$ respectively,
that is,
$$\langle \phi_i, \phi_j \rangle_\rho = \delta_{ij}$$
and
$$\langle \theta_i, \theta_j \rangle_\sigma
=
\delta_{ij}.$$

We will see that these restrictions lead to the product polynomials
$\psi_{i,j}(x,y) = \phi_i(x) \theta_j(y)$.

\subsection{Proof}

Since $\basisFuncSymbol_{i,j} (x, y)$ has maximum degree of
$i$ in $x$ and $j$ in $y$,
it can be written as
\begin{equation}
    \label{eqn:desiredBasisInProduct}
    \basisFuncSymbol_{i,j}(x,y)
    =
    \sum_{p\le i, q\le j}
    A_{ijpq}\phi_p(x) \theta_q(y)
\end{equation}
Substituting \eqnRef{desiredBasisInProduct} in \eqnRef{innerIMJN}
we get
\begin{equation}
    \label{eqn:reductionByOrthogonality}
    \begin{array}{rcl}
    \delta_{im}\delta_{jn}
    &=&
    \sum_{p\le i, q\le j}
    \sum_{r\le m, s\le n}
    A_{ijpq} A_{mnrs}
    \langle \phi_p(x), \phi_r(x) \rangle_\rho
    \langle \theta_q(y), \theta_s(y) \rangle_\sigma
    \\
    &=&
    \sum_{p\le i, q\le j}
    \sum_{r\le m, s\le n}
    A_{ijpq} A_{mnrs}
    \delta_{pr} \delta_{qs}
    \\
    &=&
    \sum_{p\le\min(i,m), q\le\min(j,n)}
    A_{ijpq} A_{mnpq}
    \end{array}
\end{equation}

We prove the result by induction on $k=m+n$ showing that
$$A_{ijmn}=\delta_{im}\delta_{jn}.$$
For $k=0$ from \eqnRef{reductionByOrthogonality}
we have immediately that
$A^2_{0000}=1$ and $A_{ij00}A_{0000}=0$
for $i+j>0$.
Since the signs of the one-dimensional orthogonal polynomials
are arbitrary,
we can take $A_{0000}=1$.
The second relations implies that $A_{ij00}=0$
for
$i+j>0$.
This establishes the result for $k=0$.

Assuming that it is true for $k-1$,
we examine \eqnRef{reductionByOrthogonality} with $m+n=k$.
Setting $i=m$ and $j=n$ in \eqnRef{reductionByOrthogonality}
we have $A^2_{mnmn}=1$
allowing us to choose $A_{mnmn}=1$.
Then, for any $(i,j) \neq (m,n)$
we have $A_{ijmn}A_{mnmn}=0$
implying that $A_{ijmn}=0$,
thus establishing the result.

\section{Accuracy of Integrators using a Chord Slope rather than the True Derivative.}
\label{sec:appendixRKConv2}

In this paper we used fine-scale simulation to estimate the derivatives of the
coarse variables from their chord slopes.
We then used the chord slope in
various numerical tasks such as finding steady states and integration.
In this appendix we address the impact of the errors due to using chord
slopes on integrators.
Note that this is unrelated to any errors due to switching back and forth 
between fine and coarse variables which must be analyzed separately.

We have used both fixed step-size methods and off-the-shelf codes that 
automatically pick the step size to control an error estimate.
We show below the following two lemmas that are valid for Runge-Kutta (RK)
and multi-step methods (and probably valid for other classes of methods,
but each needs to be examined individually):
    (1) When a fixed step-size method of order $p$ is used,
    if the chord is of length ${\inTstep} = \mathcal O({\outTstep}^p)$
    then the resulting method continues to be order $p$;
    (2) When an automatic method is used, if the error in the chord estimate
    of the derivative is bounded by the error control amount, 
    then the one-step error of the resulting method is generally increased
    by no more than a constant factor.
We discuss these results for the equation $\dot y = f(y)$.

\textbf{Fixed step-size methods}.  Any method---whether RK or multi-step---advances from one step to the next
by computing an approximation to $y(t+{\outTstep}) - y(t)$ from a linear combination 
of approximations to the derivatives at various points---%
all multiplied by ${\outTstep}$ (and also a linear combination of past values in the case of multi-step methods). 
If the method has order $p$ it means that the approximation to the change
from $t$ to $t+{\outTstep}$ has an error of ${\rm O}({\outTstep}^{p+1})$.  
Using a chord slope approximation to a derivative over distance ${\inTstep}$ 
gives an error in the derivative estimate of ${\rm O}({\inTstep})$, 
so the error in the integration formula is ${\rm O}({\inTstep}{\outTstep})$.  
Hence, if ${\inTstep} = {\rm O}({\outTstep}^p)$ then the additional error is ${\rm O}({\outTstep}^{p+1})$,
so the formula remains of order $p$.

If the method handles stiff equations and hence needs a Jacobian 
to solve the nonlinear implicit equation, as long as the Jacobian is calculated numerically 
(which is the case in most automatic codes), this result does not change, 
as now the non-linear equation to be solved involves the chord slope 
rather than the derivative.

\textbf{Automatic methods.} Automatic method adjust the step size (and possibly also the order)
from step to step to control an error estimate.  
In the early days of automatic integrators, there was a lot of discussion about 
whether to control the error per step, or the error per unit step.  
In error-per-unit-step, the error estimate for a step is controlled 
to be less than ${\outTstep}\epsilon$ where ${\outTstep}$ was the step size used and $\epsilon$ 
was the desired error. The reasoning behind this approach was that, 
if the errors accumulated more or less linearly from step to step 
(that depends very much on the stability or otherwise of the differential equation) 
then the error at the integration end point should be roughly proportional to the integration interval times $\epsilon$.  
However, if the system is stiff, errors are strongly damped from step to step, 
so the dominant error is the error in the last step, around $\epsilon$.  
Many automatic codes, having gone to the effort to estimate the error in a step, 
correct the solution by the error estimate (although there is now no direct estimate
of the higher-order error, an assumption 
that higher derivatives behave similarly to lower derivatives can provide some comfort).  
An interesting aspect of this process is that if the original control was on error per step, 
the error control now becomes per-unit-step because of the increased order.

Regardless of the actual control mechanism used,
all methods add a linear combination of derivative estimates multiplied by ${\outTstep}$
(and possibly a combination of past solution values)
so that the influence of the slope estimates is of the form
${\outTstep} \sum_{i=0}^k \beta_i s_i$.
If each $s_i$ is controlled to have an error of no greater than $\epsilon$, 
the contribution to the error is no worse than $A{\outTstep}\epsilon$, where $A = \sum_{i=0}^k |\beta_i|$.  
Thus the one step error is no worse than $A+1$ larger than before.

Most automatic codes provide both a relative error tolerance and an absolute error tolerance.  
Generally we are interested in errors relative to the size of the solution.  %
However, when the solution is almost zero it may be impossible to get an error 
which is small relative to the solution, 
so the reason for an absolute error control is to defuse that problem.  
If either relative or absolute error is bounded, the step proceeds.

\begin{figure}[ht]
\centering
\includegraphics[width=0.5\textwidth]{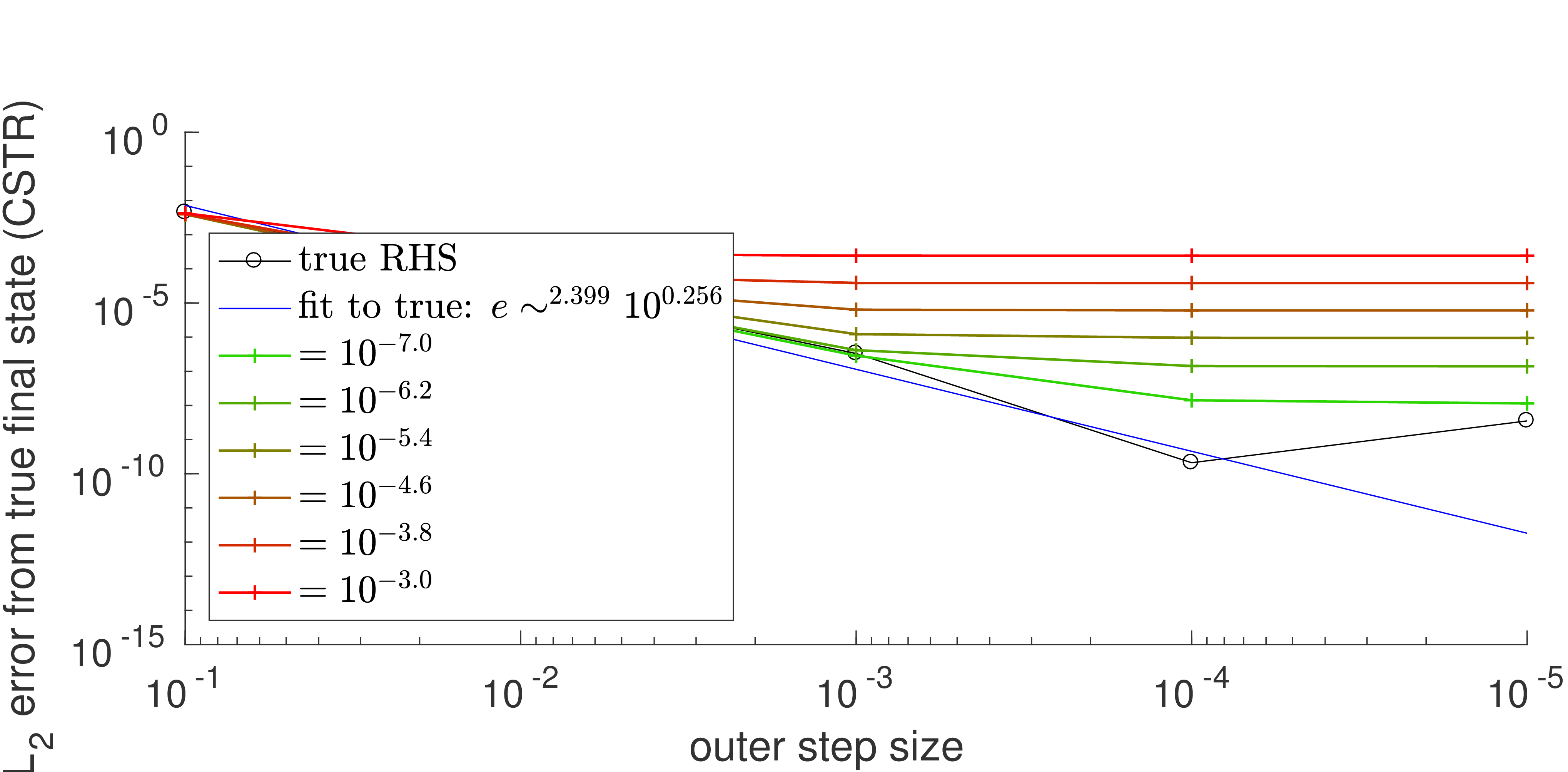}
\caption{
    \textbf{
        As the inner step size ${\inTstep}$ decreases,
        the error (compared to direct integration) of a {\em projective} integration
        becomes bounded instead by the integrator's intrinsic (outer) step size ${\outTstep}$.
    }
    Projective integration is performed
    on a system of two ODEs modeling a single reversible reaction in a
    continuously-stirred tank reactor with in- and out-flow.
    Similar to \figRef{hTauPI}, we
    quantify the difference
    between direct integration and projective integration,
    for various values of ${\inTstep}$ and ${\outTstep}$.
    The true solution at $t=\dhdtDeltaTVal$ was found by integrating
    using MATLAB's \texttt{ode45} with an
    absolute tolerance of $10^{\dhdtatolLogTen}$ and a 
    relative tolerance of $10^{\dhdtrtolLogTen}$
    The series of black circles give the error at $t=\dhdtDeltaTVal$
    that results from using integration using the true RHS function \eqnRef{trueRHS}
    in an explicit second-order Runge-Kutta integration scheme
    of (outer) step size ${\outTstep}$.
    The colored curves use the same integrator and outer step size,
    but approximate the RHS function with a difference map
    analogous to the coarse difference map of \eqnRef{coarseDifferenceMap}.
    Error is evaluated by taking the 2-norm of the vector difference
    between the projective integration solution $\vec x(\dhdtDeltaTVal)$
    and the true solution.
}
\label{fig:hTau2D}
\end{figure}

We demonstrate the convergence of error as ${\outTstep}$ is decreased for various ${\inTstep}$
on a two-ODE continuously-stirred tank reactor (CSTR) problem with in- and out-flow
in \figRef{hTau2D}.
``Full'' integration is by \texttt{ode45} with the default tolerances,
and the projective integration is a second-order Runge-Kutta scheme.
For this example, we model the reaction
\begin{equation}
\begin{array}{rcl}
&& A \xrightleftharpoons[b]{a} B \\
\frac{\diff}{\diff t}
\vec x
&=&
\mat J
\cdot
\vec x
+V
\left[ \begin{array}{c} x_{A,\text{in}} \\ x_{B,\text{in}} \end{array} \right],
\end{array}
\end{equation}
where $\vec x = \left[ \begin{array}{c} x_A \\ x_B \end{array} \right]$ is the system state,
and
$\mat J = \left[ \begin{array}{cc} -a-V & b \\ a & -b-V \end{array} \right]$ is the Jacobian.
The system parameters are
the reaction rates $a=\cstraVal$ and $b=\cstrbVal$,
the input flow rate $V=\cstrVVal$,
and
the input concentrations $x_{A,\text{in}}=\cstrxAinVal$ and $x_{B,\text{in}}=\cstrxBinVal$.
The eigenvalues of $\mat J$ are
$\lambda_1 = -(a+b+V)$ and $\lambda_2=-V$.
We first take the initial condition to be $\vec x=\left[\begin{array}{c}\cstrxAInitVal \\ \cstrxBInitVal\end{array}\right]$,
integrate until $t=-1/\lambda_1/2$, half of the first timescale.
We use this state as the common initial condition for the integrators in \figRef{hTau2D}.
As can be seen, the error of the method decreases with the inner step size $\inTstep$
until a bound governed by the outer step size $\outTstep$ is reached.
As predicted by the analysis above, this governing function is close
to the curve $\rm O(\outTstep^p)$.

\setcounter{subfigure}{0}

\end{document}